\documentclass[
reprint,
superscriptaddress,
amsmath,
amssymb,
aps,
longbibliography,
prx
]{revtex4-2}
\usepackage{amsmath}
\usepackage{mathtools}
\usepackage{soul}
\usepackage{amssymb}
\usepackage{graphicx}
\usepackage{bm}
\usepackage[colorlinks, linkcolor=blue, citecolor=blue, urlcolor=blue,breaklinks=true]{hyperref}
\usepackage{color,dsfont,multirow,booktabs}
\usepackage{braket}
\usepackage{tabularx}
\usepackage{lipsum,booktabs}

\begin{document}

\title{Advances in Quantum Radar and Quantum LiDAR}

\author{Ricardo Gallego Torrom\'e}
\email{rigato39@gmail.com}
\affiliation{Department of Mathematics Faculty of Mathematics, Natural Sciences and Information Technologies University of Primorska, Koper, Slovenia}
 
\author{Shabir Barzanjeh}
\email{shabir.barzanjeh@ucalgary.ca}
\affiliation{Institute for Quantum Science and Technology, and Department of Physics and Astronomy University of Calgary,
2500 University Drive NW, Calgary, Alberta T2N 1N4, Canada}

\begin{abstract}
Quantum sensing, built upon fundamental quantum phenomena like entanglement and squeezing, is revolutionizing precision and sensitivity across diverse domains, including quantum metrology and imaging. Its impact is now stretching into radar and LiDAR applications, giving rise to the concept of quantum radar. Unlike traditional radar systems relying on classical electromagnetic, quantum radar harnesses the potential of the quantum properties of photon states like entanglement and quantum superposition to transcend established boundaries in sensitivity and accuracy. This comprehensive review embarks on an exploration of quantum radar and quantum LiDAR, guided by two primary objectives: enhancing sensitivity through quantum resources and refining accuracy in target detection and range estimation through quantum techniques. We initiate our exploration with a thorough analysis of the fundamental principles of quantum radar, which includes an evaluation of quantum illumination protocols, receiver designs, and their associated methodologies. This investigation spans across both microwave and optical domains, providing us with insights into various experimental demonstrations and the existing technological limitations.
Additionally, we review the applications of quantum radar protocols for enhanced accuracy in target range determination and estimation. This section of our review involves a comprehensive analysis of quantum illumination, quantum interferometry radar, and other quantum radar protocols, providing insights into their contributions to the field. This review offers valuable insights into the current state of quantum radar, providing a deep understanding of key concepts, experiments, and the evolving landscape of this dynamic and promising field.
\end{abstract}

\maketitle
\tableofcontents
\section{Introduction}

Quantum sensing harnesses inherent quantum phenomena such as entanglement, quantum interference, and quantum squeezing to surpass conventional boundaries of accuracy and sensitivity. It finds applications in quantum metrology, quantum imaging, and various facets of quantum technology, enabling the achievement of unprecedented levels of precision and sensitivity previously considered unattainable. The implications of quantum sensing span the entire spectrum of scientific exploration, encompassing both foundational research and practical real-world applications. The profound influence of quantum sensing is supported by a substantial body of scientific literature, demonstrating its significance in the field \cite{Degen Reinhard Cappellaro, Pirandola, Zheng, Aslam, Harris 2009, Mayte}.

Within the expansive landscape of quantum sensing, a burgeoning field that intersects with radar and Light Detection and Ranging (LiDAR) applications has emerged, known as quantum radar. Traditional radar systems rely on classical electromagnetic waves to detect and locate objects by measuring the signals they reflect \cite{Skolnik, Slepyan, WAILOKLAI201858}. The primary objective of quantum radar, however, is to exploit the unique quantum properties offered by entanglement or nonclassicality to enhance target detection capabilities. Quantum radar operates by sending entangled microwave or optical photons toward the target and then measuring the correlation between partner photons and the returned signal. The initial quantum entanglement (non-classical properties of the states) enables quantum radar to surpass classical radar systems in terms of sensitivity and accuracy. Although the concept of quantum radar has been discussed for some time \cite{Marco Lanzagorta 2011, Pirandola, Shapiro2019, Gallego Torrome Bekhtil Knott 2021}, recent years have witnessed a surge in interest and progress, drawing the attention of a broader audience. Given the dynamic and evolving nature of this field, it becomes imperative to provide a concise overview of the current state of the art and its future prospects.

This comprehensive review aims to shed light on the recent advancements within the fields of quantum radar and quantum LiDAR. The discussion is structured along two primary lines of inquiry: the utilization of quantum resources to improve sensitivity and the use of quantum techniques to enhance accuracy in target range determination and the estimation of other target-related parameters.
Starting with an exploration of the foundational concepts of quantum radar, we then review different quantum illumination protocols and receiver designs, all the while conducting an in-depth examination of related quantum radar protocols. It is of utmost importance to gain a clear understanding of the current state of experimental demonstrations in this field and to grasp the limitations of applicability in both microwave and optical (LiDAR) frequency domains, as well as to elucidate the genuine constraints faced by quantum radar technology. Subsequently, we shift our focus to quantum radar protocols aimed at enhancing accuracy. We begin with an exploration of recent developments in quantum illumination for this purpose and then venture into other protocols such as quantum interferometry radar and the Maccone-Ren proposal introduced in \cite{MacconeRen2019} and further developed recently in \cite{MacconeZhengRen}.

While the objective of this review is to provide the reader with insights into the current state of the art, our discussion has been intentionally kept general. We highlight key concepts and experiments that illustrate the aforementioned points. Quantum radar, as a field, is still in its infancy, and it would be unfair to judge it solely by its current limitations. Particularly given the evolving nature of this field, it has become apparent that there is still much to learn and understand about the mechanisms that can be harnessed in quantum radar. Therefore, a balanced perspective is essential, recognizing both its current challenges and its immense potential for future advancements. 

\begin{figure*}[t]
\begin{center}
 \includegraphics [width=1\linewidth]{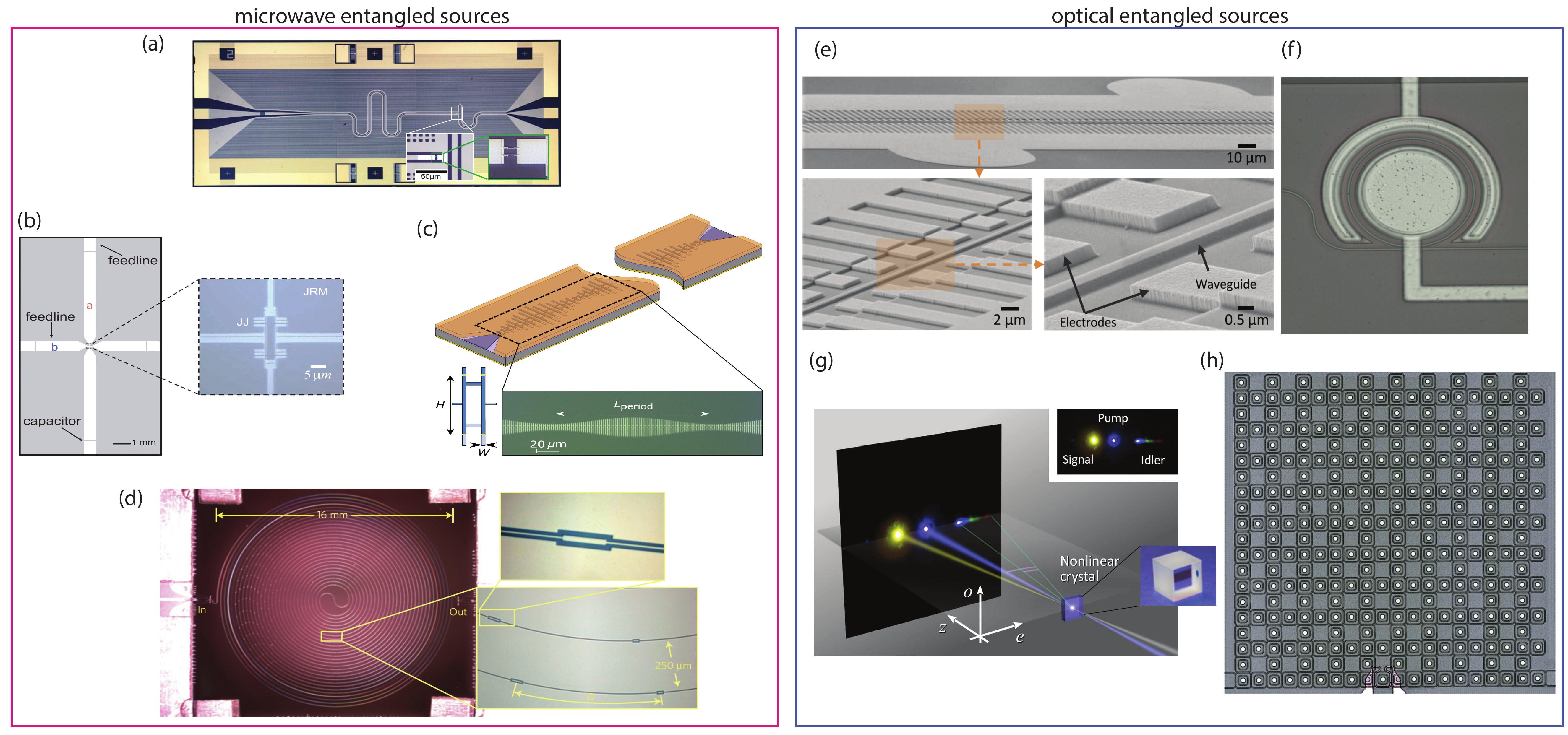}
\caption{ Different sources of entanglement and nonclassicality in microwave and optical frequency ranges. 
(a) Single junction parametric amplifier (JPA) \cite{Zhong2013}. (b) Josephson ring modulator (JRM) containing four Josephson junctions \cite{Abdo}. (c) A traveling wave amplifier (TWPA) fabricated based on an array of Josephson junctions \cite{Planat2020}. (d) Kinetic inductance traveling wave amplifier \cite{EOM}. (e) Entangled photons on a nanophotonic chip \cite{Usman}. (f) entangled pair generation in Silicon micro-ring resonator \cite{Savanier}. (g) Entanglement generation using a nonlinear crystal \cite{Manzoni}. (h) Robust entanglement generation in topological photonic insulators \cite{Afzal}.}
\label{figgeneral}
\end{center}
\end{figure*}

\section{Notion of quantum radar}

\subsection{Quantum radar within the context of quantum sensing}
Quantum radar or LiDAR can be viewed as a subset of quantum sensing, primarily focused on extracting classical parameters from conventional objects, such as speed, target range, and more. Quantum sensors, in a broader context, can be divided into two distinct scenarios: \textbf{Category (1)}, where the probing source utilizes quantum radiation and the detector excels at measuring quantum radiation reflected from the sample/target. In contrast, \textbf{Category (2)} employs classical light for probing, while the measurement or detection process operates within the quantum regime. Each of these categories of quantum sensors has its advantages and disadvantages. Category 1, for instance, is useful for improving measurement sensitivity, such as quantum illumination, in a non-invasive and stealth fashion, thanks to the low power of quantum radiation. Nevertheless, it faces difficulties in long-range applications where noise and losses can significantly surpass the quantum properties, potentially resulting in the degradation or breaking of nonclassicality or entanglement. On the other hand, Category 2 can find applications in long-range detection (such as LiDAR) by employing extremely sensitive detectors or chains of entangled detectors \cite{PhysRevLett.124.150502}. Nevertheless, this approach may entail the risk of causing harm to the samples and typically demands substantial power for extended-range operations. The sensitivity of Category 2 sensors in some cases can still remain comparable to that of classical sensors, as they do not rely on classical radiation for probing the sample.

Within the quantum sensing community, alternative classification schemes are prevalent. For instance, in Ref. \cite{Harris 2009}, quantum sensors are divided into three discernible types, predicated on their unique utilization and the characteristics of entanglement and other related non-classical phenomena: 
\begin{itemize}
\item {\bf Type 1 quantum sensor}: A Type-1 sensor transmits a non-classical state that is not
entangled with the receiver. 
\item {\bf Type 2 quantum sensors}: A Type-2 sensor transmits a classical state that is not entangled
with the receiver and employs a non-standard (quantum-enhanced) detection procedure in that receiver.
\item {\bf Type 3 quantum sensors}: A Type-3 sensor transmits a state — which may be classical or
non-classical — that is entangled with the receiver.
\end{itemize}
Maccone-Ren's quantum radar \cite{MacconeRen2019} is a quantum protocol of Type 1, while quantum interferometric radar and quantum illumination are protocols of Type 3. Examples of Type 2 can be found in \cite{Harris 2009}.

We adopt the concept of quantum radar and LiDAR as a quantum sensing protocol that employs non-classical quantum resources, whether in the source, detector, or both, to achieve greater accuracy and sensitivity in target detection tasks compared to equivalent classical illumination systems. Consequently, the scope of quantum radar extends beyond quantum radar protocols based on quantum illumination. Nonetheless, protocols rooted in quantum illumination have played a substantial role in advancing quantum radar research in recent years, as evidenced by references \cite{Marco Lanzagorta 2011,Shapiro2019, Gallego Torrome Bekhtil Knott 2021}.

This review article aims to provide an overview of recent developments in the field of quantum radar, encompassing enhancements in both sensitivity and accuracy for determining target range. We acknowledge that our perspective may be somewhat constrained when compared to what is occasionally termed "quantum radar/LiDAR" in the literature. In certain cases, the term "quantum" is employed to describe a protocol or device that, in one manner or another, exploits the quantum properties of electromagnetic fields and matter. While the definition of radar could encompass the optical spectrum, in this context, we maintain the concept of quantum radar within the microwave domain and reserve the term LiDAR exclusively for the optical frequency range.

\subsection{Nonclassical sources: entanglement and squeezing}
Quantum radar/LiDAR protocols harness the distinct properties of electromagnetic radiation, specifically exploiting quantum phenomena like quantum entanglement, quantum squeezing, and various other forms of quantum properties, to enhance their effectiveness in target detection applications. Depending on the wavelength employed, several methods are available to generate entanglement and squeezing, with applications in quantum sensing. Figure \ref{figgeneral} illustrates various devices in both optical and microwave domains that produce nonclassical radiation, predominantly entangled pairs and squeezing. The common point among these devices, irrespective of their wavelength, is their capacity to generate nonclassical light. A general approach employed in many of these devices involves parametric amplification, whether degenerate (phase-sensitive) or non-degenerate (phase-insensitive), to produce nonclassical radiation. The output of a non-degenerate amplifier can be characterized by the signal $a_S$ and idler $a_I$ modes, which exhibit a certain relationship with the input field operators $a_0$ and $a_0'$ \cite{Caves}
\begin{eqnarray}\label{TMW}
    a_S&=&\sqrt{G}a_0+e^{i\phi}\sqrt{G-1}a_0'^\dagger,\nonumber\\
    a_I&=&\sqrt{G}a_0'+e^{i\phi}\sqrt{G-1}a_0^\dagger
\end{eqnarray}
where $G$ is the gain of the amplifier and $\phi$ is the phase. 
Note that the existence of the second term ($\propto \sqrt{G-1}$) in each line of Eq. (\ref{TMW}) is required to fulfill the bosonic commutation relation $[a_{S/I},a_{S/I}^\dagger]= 1$. The non-degenerate amplification (under specific conditions) can result in the generation of entanglement or two-mode squeezing, which arises as a direct consequence of the correlation between the idler and signal
\begin{equation}\label{correlation}
   C_{q0}= \langle a_Sa_I\rangle=\sqrt{N_S(N_S+1)},
\end{equation}
considering $N_S=\langle a_S^\dagger a_S\rangle=\langle a_I^\dagger a_I\rangle=G-1$. In general, the output of a non-degenerate amplifier becomes entangled if the parameter \cite{Barzanjeh et al.}
\begin{equation}
    \epsilon_\text{ent}=\frac{|C_{q0}|}{\sqrt{N_S N_I}},
\end{equation}
exceeds one $\epsilon_\text{ent}>1$, where  $N_{S/I}=\langle a_{S/I}^\dagger a_{S/I}\rangle$. In this case, the cross-correlation between the idler and signal exceeds the number of photons in each mode. As we will see, this cross-correlation plays a central role in Gaussian quantum illumination and aids in distinguishing the presence and absence of a target. In a specific scenario, when both input field operators are identical (correlated), we can  (informally) treat $a_0= a_0'$ which results in degenerate amplification 
\begin{equation}
    a_S=a_I=\sqrt{G}a_0+e^{i\phi}\sqrt{G-1}a_0^\dagger.
\end{equation}
It can be demonstrated that degenerate amplification under the right phase conditions can result in single-mode vacuum squeezing.

In the microwave domain, entanglement generation typically relies on strong nonlinearity achieved through Josephson junctions, known as Josephson Parametric Amplifiers (JPAs). Within this context, various designs are feasible, including single junctions \cite{Zhong2013}, Josephson ring modulators (JRM) \cite{Abdo}, and traveling wave amplifiers based on Josephson junction \cite{Planat2020} or kinetic inductance superconductors \cite{EOM}. 

In contrast to the microwave domain, optical systems face limitations in the strength of materials' nonlinearity, consequently restricting the degree of squeezing or amplification achievable. To overcome this limitation, nano \cite{Usman} or micro-structures such as micro-ring resonators \cite{Savanier} or topological systems \cite{Afzal} can be employed. These nano or micro-fabricated structures prolong the duration of interaction, facilitating the production of broadband and bright entangled pairs or squeezing, even when weak nonlinearity is present. The general techniques for creating optical entanglement predominantly employ Spontaneous Parametric Down-Conversion (SPDS) \cite{Manzoni}, often utilizing the Kerr nonlinearity in the materials.

\section{The principle and theory of Quantum illumination radar}
The foundational principles for enhancing sensitivity in quantum illumination protocols, as opposed to their classical counterparts, are firmly grounded in the framework of quantum detection and estimation theory introduced by Helstrom \cite{Helstrom1976}. His theory is based on quantum hypothesis testing by deducing the optimal measurement operator aimed at minimizing the error probability associated with the single-copy discrimination of quantum states. This theory represents an extension and generalization of Chernoff's theory \cite{Chernoff} and holds particular relevance within the domain of quantum discrimination. 
When dealing with particular types of entanglement-breaking channels, the common practice involves applying the quantum Chernoff's theorem, as extensively utilized in the exploration of sensitivity enhancement within the context of quantum illumination \cite{Audenaert et al.}. We won't go deeper into this topic, but we recommend interested readers to refer to this review paper for more information \cite{Weedbrook12}.

 \begin{figure}[t]
\begin{center}
 \includegraphics [width=1\linewidth]{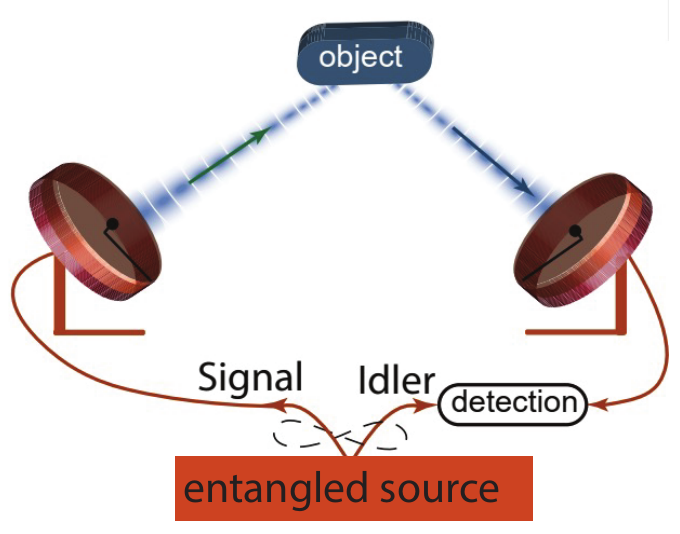}
\caption{ Quantum Illumination General Framework. This protocol employs an entangled source to create entangled idler and signal beams. While the idler remains preserved, the signal is utilized to investigate an area suspected to contain a target. Through a joint measurement involving the retained idler and the returned signal, we acquire information regarding the absence or presence of the target. }
\label{fig: quantum illumination}
\end{center}
\end{figure}

\label{Quantum Illumination}
\subsection{The concept of Quantum illumination}

In the general context of quantum illumination, a pair of entangled photon beams is generated. One of these beams serves as the signal, while the other remains as the idler, preserved throughout the entire experiment until a joint measurement of the signal/idler system is conducted. The signal beam is deployed to probe the region where a potential target may be located and subsequently detected upon its return as a received signal beam. Typically, the comparison between the received signal and idler beams involves joint measurements of their in-phase and quadrature components, although other forms of joint measurements, such as direct photon counting detection methods, are currently employed as well.  Fig. \ref{fig: quantum illumination} depicts schematically the generic protocol of quantum illumination. The quantum illumination theoretically offers higher sensitivity and/or an improved signal-to-noise ratio (SNR) compared to the utilization of classical light beams possessing equivalent brightness and energy characteristics. This advantage becomes especially evident in scenarios characterized by low-intensity signals, low target reflectivity, and the presence of environmental noise \cite{Lloyd2008, Tan, Barzanjeh et al.}. 

The enhancement achieved through quantum illumination remains robust in the face of noise and losses. This enhancement is a consequence of the residual correlations, in Eq. (\ref{correlation}), inherited from the initial entanglement, setting it apart from classical sources. Typically, the analysis of quantum illumination encompasses a range of conditions, including but not limited to the following Quantum Illumination Assumptions (QIA):
 \begin{itemize}

\item The signal-idler radiation consists of an extremely small number of photons per mode, where $N_S\ll 1$.

\item There is a large time-bandwidth product i.e., $M=\,T W\,\gg 1$.

\item The reflectivity index in the presence of a target is very low, with $0<\eta\,\ll 1$ and $\eta =0$ in the absence of the target.

\item  The average photon number per mode for background noise is very large $N_B\gg 1$.

\end{itemize}

\subsection{Single-photon quantum illumination}
The concept of quantum illumination was initiated by the pioneering work of Lloyd \cite{Lloyd2008}. He studied the advantages of incorporating entanglement in the detection of individual photons, particularly in situations characterized by challenging conditions such as losses and noise. Lloyd's work emphasized an improvement in performance compared to the use of single-photon states lacking entanglement. Lloyd's quantum illumination protocol held theoretical significance as it demonstrated how the benefits stemming from quantum correlations rooted in entanglement could persist even in the presence of losses within a moderately noisy setting.









Expanding upon this foundational work, Shapiro and Lloyd \cite{ShapiroLloyd} further solidified this concept by affirming that, under ideal circumstances, Lloyd's "quantum illumination" system achieves performance at most comparable to that of a coherent-state transmitter. Importantly, this system has the potential to provide substantial performance improvements when operating beyond the confines of the single-photon regime. These analyses were based on two different regimes of interest, {\it good regime} and {\it bad regime} \cite{ShapiroLloyd}. For the single-photon state beam protocol, the {\it good regime} happens when $\eta/ N_B \,> 1$, while for beams formed by entangled photons, the {\it good regime} happens when $\eta > N_B/M$.  For non-entangled single-photon states, the error probability  $Pr^+(e)_{SP}$ is bounded by the Chernoff bound
\begin{align}
Pr^+(e)_{SP}=\,e^{-N\eta}/2,\quad \textrm{for}\,\, \eta\gg N_B,
\label{Lloyd good regime SP}
\end{align}
For quantum illumination, the probability of error is bounded as
\begin{align}
Pr^+(e)_{QI}=\,e^{-N\eta}/2,\quad \textrm{for}\,\, \eta\gg N_B/M,
\label{Lloyd good regime QI}
\end{align}
showing an enhancement of the region of validity of the good regime (where the probability of false positive is very small) in the case of quantum illumination, despite that there is not an enhancement in the probability of false positive.

In the so-called {\it bad regime}, when $\eta /N_B<1$, we have the following probabilities of false detection,
\begin{align}
Pr^-(e)_{SP}=\,e^{-N\eta^2/8N_B}/2,\quad \textrm{for}\,\, \eta\ll N_B,
\label{Lloyd bad regime SP}
\end{align}
for non-entangled signals and
\begin{align}
Pr^-(e)_{QI}=\,e^{-N\eta^2 M/8N_B}/2,\quad \textrm{for}\,\, \eta\ll N_B/M
\label{Lloyd bad regime QI}
\end{align}
when there is entanglement.
Thus for a low-reflective systems, the probability of error in quantum illumination is reduced drastically with $M\gg 1$ with respect to single photon states beams. Furthermore, there is an enhancement of the region of validity of this result, from $\eta\ll N_B$ to $\eta\ll N_B/M$.

\subsection{Gaussian quantum illumination}

\begin{figure}[t]
\begin{center}
 \includegraphics [width=0.8\linewidth]{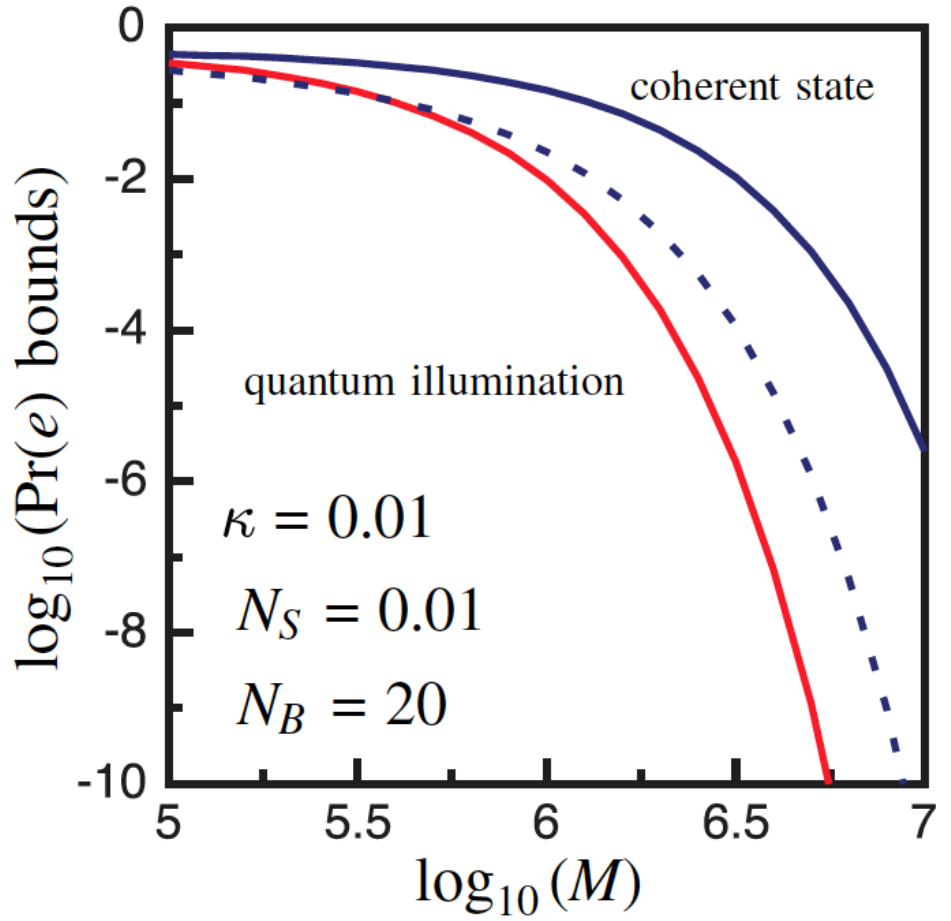}
\caption{ Advantage of Gaussian quantum illumination with respect to coherent light illumination \cite{Tan}. Here. $\eta=\kappa$ is the reflectivity of the target, $N_s$ shows the number of photon of the signal, and $N_B$ represents the background noise.}
\label{fig:advantagegqi}
\end{center}
\end{figure}

The seminal paper of Tan et al.  \cite{Tan}  demonstrated that quantum illumination employing Gaussian states offers distinct theoretical advantages compared to all classical illumination methods, including those based on coherent light, as illustrated in Figure \ref{fig:advantagegqi}.

Under the QIA,  for coherent light illumination the error probability $Pr(e)$ is given by the quantum Chernoff bound
\begin{align}
Pr(e)_{CI}\leq \, e^{-M\eta N_S/4N_B}/2.
\label{coherent illumination 2}
\end{align}
In contrast, Gaussian quantum illumination results in
\begin{align}
Pr(e)_{QI}\leq \, e^{-M\eta N_S/N_B}/2,
\label{quantum illumination 2}
\end{align}
when $N_B\gg 1 $, $0<\eta \ll 1$ and $N_S\ll 1$, highlighting a fourfold enhancement in error probability or equivalently, a 6 dB improvement in signal-to-noise ratio (SNR) \cite{Di Candia et al., de Palma Borregaard, Pirandola Lloyd}.

\subsection{Different Receivers for Gaussian quantum illumination}

Designing a receiver for quantum illumination that can harness the full 6 dB improvement in SNR predicted by the comprehensive theoretical model is challenging. In the context of Gaussian quantum illumination, the observable signifying the presence of a target is given by
 \begin{align*}
 \widehat{O}_{RI}(m):=\hat{a}_{Rm}\,\hat{a}_{Im}
 \end{align*}
  namely, the observable is the expectation value of phase-sensitive cross-correlation $\langle \hat{a}_{Rm}\,\hat{a}_{Im}\rangle$
where
\begin{align*}
\hat{a}_{Rm}=\,\sqrt{\eta}\,\hat{a}_{Sm}+\sqrt{1-\eta}\,\hat{a}_{Bm}
\end{align*}
is the annihilation operator of the received mode $m$, while $\hat{a}_{Im}$ stands for the annihilation operator of the idler mode $m$. Additionally, $\hat{a}_{Bm}$ is the operator for the background field noise and $\hat{a}_{Sm}$ corresponds to the operator for the signal mode. We consider the hypothesis that no target is present as $H_0$, whereas $H_1$ represents the hypothesis of the presence of a target. The outcome takes the following form:
\begin{eqnarray}
C_q&=&\langle \hat{a}_{Rm}\,\hat{a}_{Im}\rangle_{H_1}=\,\sqrt{\eta\,N_S\,(N_S+1)},\nonumber\\
C_q&=&\langle \hat{a}_{Rm}\,\hat{a}_{Im}\rangle_{H_0}=\,0.
\end{eqnarray}
The first equation is the reflectivity of the object multiplied by the cross-correlation in Eq. (\ref{correlation}). Note that for classical light illumination, $C_c=\langle \hat{a}_{Rm}\,\hat{a}_{Im}\rangle_{H_1}=\,\sqrt{\eta}\,N_S$. Therefore, in the limit $N_S<<1$, the value of the quantum correlation exceeds the classical correlation  $C_q>C_c$.
Nonetheless, it is not feasible to directly measure the operator $\widehat{O}_{RI}(m)$ in experiments. This operator can be expanded into individually observable quadrature operators, but the Heisenberg uncertainty relations prevent the simultaneous measurement and acquisition of knowledge about all four of them (two pertaining to the idler beam and two relating to the return beam).

\begin{figure}[t]
\begin{center}
 \includegraphics [width=1\linewidth]{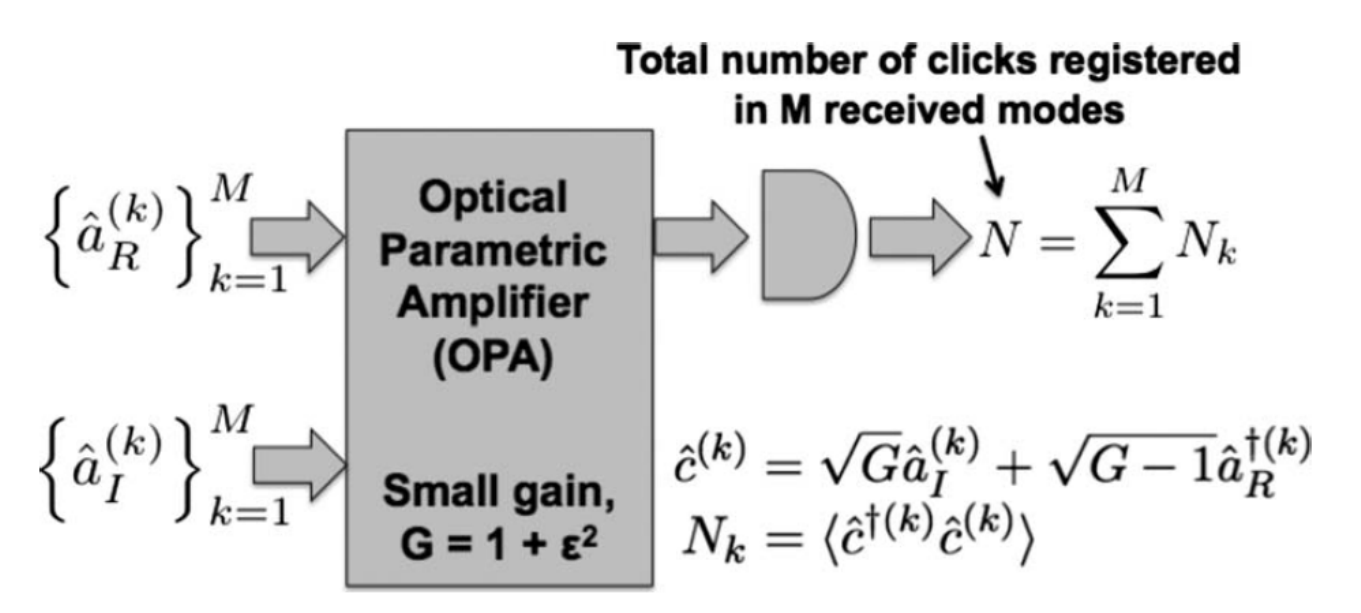}
\caption{Within the OPA receiver, both the return singal and idler modes serve as inputs into an OPA with gian $ G$. The cumulative count of photons, represented as $N$, is recorded at a single output port, as described in \cite{Guha Erkmen 2009}.}
\label{reciever}
\end{center}
\end{figure}

\textbf{Receiver based on Optical Parametric Amplification.} In the search to identify an optimal receiver for Gaussian QI, Guha and Erkmen \cite{Guha Erkmen 2009} suggested mixing the received signal and the idler beam in an Optical Parametric Amplification (OPA) and the Phase-Conjugator (PC). The OPA detector utilized for quantum illumination employs single spatial mode fields, where the signal mode with frequency $\omega_{S_0}+\delta\omega$ exhibits correlation solely with an idler mode at frequency $\omega_{I_0}-\delta \omega$, see Fig. \ref{reciever}. This condition ensures that both frequencies fall within the phase-matching bandwidth, satisfying the condition $|\delta \omega|<\pi,W$, where $W$ represents the bandwidth.

In this receiver, the observable is given by the expectation value of the number operator.
\begin{align}
\hat{N}_T =\,\sum^M_{m=1}\,\hat{a}^{out\dag}_{m}\,\hat{a}^{out}_{m},
\label{Observable Guha Erkmen}
\end{align}
where $\hat{a}^{out}_{m}= \sqrt{G}\hat{a}_{Im}+\sqrt{G-1}\hat{a}^\dag_{Rm}$ represents the outcome involving the idler and reflected signal within the OPA, as explained in Eq. (\ref{TMW}). The coefficient $G$ is the detector's gain, which can be chosen to optimize detection performance. Here, quantity $\langle \hat{N}_T\rangle_{H_i}$ with $i=0,1$ distinguishes the absence and presence of the target. The observable \eqref{Observable Guha Erkmen}, contains all cross-correlations $\langle \hat{a}_{Rm}\hat{a}_{Im}\rangle_{H_i}$, $\langle \hat{a}_{Rm}\hat{a}^\dag_{Rm}\rangle_{H_i}$, and $\langle \hat{a}^\dag_{Im}\hat{a}_{Im}\rangle_{H_i}$ and thus all of them are now observable, as the Heisenberg principle does not prohibit their measurement.

In scenarios where the OPA receiver exhibits a small gain $G\approx 1+\epsilon$ with $\epsilon \ll 1$, and in the presence of high-intensity background noise and low-intensity signals, all while maintaining low reflectivity and assuming lossless idler and precise synchronization in time and phase between the returned and idler beams, we can calculate an upper limit for the probability of error when discriminating between hypotheses $H_0$ and $H_1 $\cite{Guha Erkmen 2009}
\begin{align}
Pr(e)_{QI}\leq \,e^{-M\,\eta\,N_S/2N_B}/2 .
\label{Guha detector Chernoff bound}
\end{align}

The OPA-based receiver protocol can not fully capture the theoretical advantage of 6 dB offered by QI over coherent illumination, which represents an optimal classical scenario. Instead, it achieves a theoretical advantage of only 3 dB in OPA detection which has been further validated through other analyses employing quantum estimation techniques \cite{Sanz et al.}.

Ref \cite{Zhang et al. 2015} demonstrated the experimental realization of Gaussian quantum illumination's enhancement over a coherent light homodyne-detection scheme using the OPA detection method. In this experiment, a laser pump with a wavelength of $\lambda_p = 780$ nm was employed in an SPDC process to generate two entangled beams at wavelengths $\lambda_s = 1590$ nm and $\lambda_i = 1530 $ nm. The noise was introduced at the same wavelength as the signal beam. The reported improvement of quantum illumination over coherent light illumination is rather modest, on the order of approximately $20\%$ in SNR.

\textbf{Phase-conjugate receiver.} Guha and Erkmen's work also introduced the concept of the phase-conjugate receiver \cite{Guha Erkmen 2009}. This receiver operates within the same domain of low signal, low reflectivity, and high noise as the OPA receiver, and it achieves the same theoretically predicted 3 dB performance enhancement. The operation of the phase-conjugate receiver involves conjugating each of the $M$ reflected modes from the target region. The result of this will be that the received mode is mixed with vacuum $\hat{a}_{Vm}$ and conjugated, resulting in $\hat{a}_{Cm} =\sqrt{2}\hat{a}_{Vm}+\hat{a}^\dagger_{Rm}$. Subsequently, these modes are combined with their respective idler modes generating the modes $\hat{a}_{Xm}=(\hat{a}_{Cm}+  \hat{a}_{Im})/\sqrt{2}$ and $\hat{a_{Ym}}=(\hat{a}_{Cm}+  \hat{a}_{Im})/\sqrt{2}$, each associated with number operators $\hat{N}_X$ and $\hat{N}_Y$, respectively. The modes are subsequently assessed using a differential balance detection technique, which involves measuring the difference between $\hat{N}_X$ and $\hat{N}_Y$. This process enables the inference of SNR under both conditions, with and without the presence of the target.

{\bf Receiver based on sum frequency generation}.
 Another optimal receiver is based on the sum-frequency generation (SFG), in the regime characterized by a large number of modes and a low-intensity source \cite{Zhuang Zhang Shapiro b}.  Utilizing this, one can design a receiver with the potential to realize the anticipated 6 dB advantage as postulated in Ref \cite{Tan}.
 
 SFG receivers rely on a set of idealized assumptions that remain beyond the reach of current technology. However, the inclusion of a feedforward (FF) mechanism in the SFG receiver elevates its performance towards reaching the Helstrom bound, particularly in scenarios with low signal brightness \cite{Shapiro2019}.

\textbf{Hetero-Homodyne Receiver and Sequential Detection:} A promising recent development emerges in a novel quantum receiver protocol presented in Ref \cite{Reichert et al. 2023}. The protocol essentially includes two innovative elements: On one front, it incorporates the hetero-homodyne detection scheme, which constitutes a cascading Positive Operator-Valued Measure that eliminates the necessity for direct quantum idler-return signal measurements. The second key component of this receiver protocol involves the implementation of the Sequential Probability-Ratio Test (SPRT) \cite{Wald 1945} for assessing the probabilities of error in target detection. Employing sequential detection with the hetero-homodyne detector provides a $3$ dB quantum advantage over the optimal classical illumination, as observed in Guha and Erkmen's study on phase conjugate and parametric amplifier receivers. However, in a sequential detection QI protocol, the hetero-homodyne receiver provides a $9$ dB quantum advantage over a traditional classical radar and a $3$ dB advantage over a classical radar with sequential detection. Furthermore, the proposed receiver does not necessitate a quantum interference between the received signal and the stored idler, enhancing its adaptability for practical implementations. We note that this receiver is constructed under the premise of preserving the idler throughout the measurement process. This assumption imposes limitations on the efficiency and maximum achievable target range.

\section{Implementation of quantum illumination radar} 
\subsection{Practical limitations} Quantum illumination, whether at microwave or optical wavelengths, presents an improved sensitivity (under certain conditions) compared to most forms of non-entangled light when considering identical idler and signal intensities. However, recent years have witnessed a close examination of the mechanisms underlying quantum illumination, casting doubts on its applicability in its conventional form. Despite potential advantages in specific configurations, quantum illumination is constrained by the requirement of prior knowledge of the target range, commonly referred to as "the range problem." This limitation primarily confines the utility of quantum illumination protocols to scanning applications. Several approaches have been suggested to address this range problem, including the utilization of modulated frequency pulses within the Joint Pulse Radar (JPR) \cite{Karsa Pirandola 2021}. When combined with conventional Doppler techniques, these pulses enable the measurement of the target's relative speed. However, alternative solutions have also emerged to tackle this challenge within quantum radar. One such solution is the Maccone-Ren quantum radar protocol \cite{MacconeRen2019}, which, while distinct from quantum illumination, offers a theoretical resolution to the range problem, and we will come back to this protocol later in our discussion.

In the context of quantum illumination protocols based on Gaussian states, particularly for medium and long-range radar applications, a significant challenge arises due to the exceedingly low signal intensities required to harness the sensitivity advantages offered by quantum illumination \cite{Jonsson et al.}. This issue becomes more pronounced in the microwave quantum radar, primarily due to signal losses incurred during propagation through the atmosphere. Although less critical for short-distance applications, the core challenge remains: quantum illumination demonstrates sensitivity benefits over coherent illumination at signal intensities that are remarkably diminutive.
To maximize the exploitation of quantum properties, one must operate at energy levels akin to single photons. To illustrate this, a single photon at 6 GHz equals to a signal magnitude of approximately -114 dBm. These estimations provide upper bounds for the intensities required to leverage the quantum advantage. This inherent limitation poses a significant practical challenge, especially in scenarios where maintaining such low signal intensities is technologically challenging.

The practical implementation of quantum illumination for radar encounters several additional challenges. It necessitates a large time-bandwidth product compared to the typically narrowband generation observed in entangled photon states, particularly in the context of microwave quantum illumination. Moreover, when operating at optical frequencies (LiDAR), the presence of Rayleigh-fading targets further complicates the scenario, as discussed in \cite{ZhuangZhangShapiro}. As a result, alternative approaches for handling idler information and extending the applicability to larger target ranges have been explored \cite{Barzanjeh et al.2019,Luong et al. 2019}.

Efficient idler storage is another critical consideration. The losses associated with the idler severely limit the maximum attainable target range \cite{Barzanjeh et al.}, and the proposed solution of employing quantum memories is contingent upon their practical availability \cite{Heshami et al. 2016}.

  \begin{figure}[h]
\graphicspath{{Images/}}
\begin{center}
 \includegraphics [width=1\linewidth]{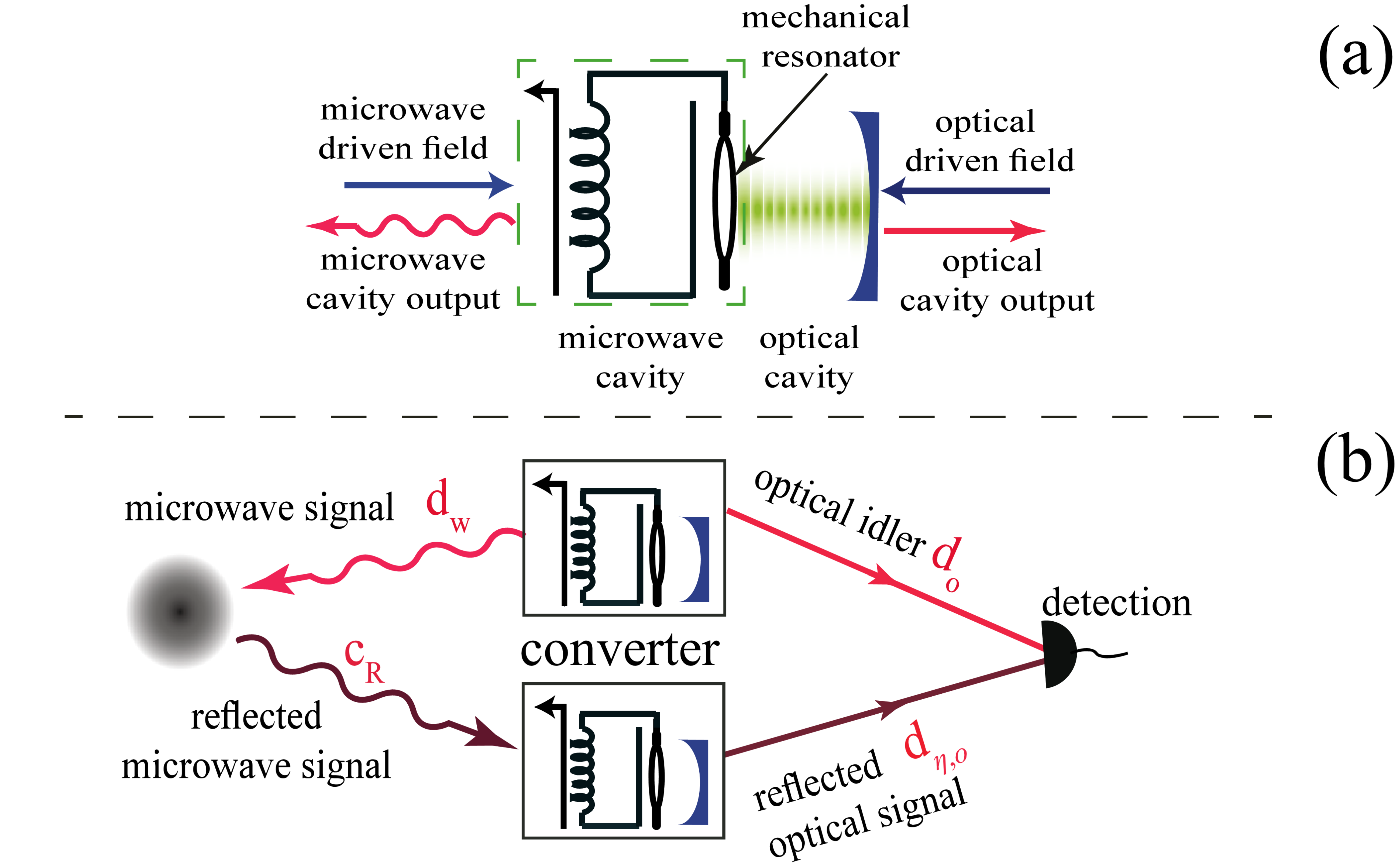}
\caption{(a) Schematic representation of an electro-opto-mechanical (EOM) converter, featuring interconnected microwave and optical cavities coupled through a mechanical resonator. (b) Application of EOM converters in microwave-optical quantum illumination. In this setup, the EOM converter generated entanglement between microwave and optical fields. Meanwhile, another EOM at the receiver's end converts the incoming microwave field into the optical domain while simultaneously executing a phase-conjugate operation \cite{Barzanjeh et al.}.}
\label{Microwaveqi}
\end{center}
\end{figure}

\subsection{Microwave quantum illumination}
The principles of quantum illumination we discussed earlier, along with the conditions for sensitivity improvement, are generally applicable across a wide range of frequencies. However, it's important to note that conventional radar systems typically function within the microwave frequency range. Microwave frequencies typically cover the range of 100 MHz to 36 GHz in standard radar applications, and in some cases, millimeter-wave radar systems can extend this range to several hundred Gigahertz (GHz) \cite{Skolnik}. This preference for microwave frequencies is primarily due to the reduced attenuation of microwave signals in the atmosphere compared to shorter wavelength regimes, making it an attractive domain for quantum illumination investigations.

The introduction of quantum illumination in the microwave domain was pioneered by Barzanjeh et al. in 2015 \cite{Barzanjeh et al.}. In this approach, an electro-optomechanical (EOM) cavity (see Fig. \ref{Microwaveqi}a) \cite{PhysRevLett.109.130503, PhysRevA.84.042342, Lauk_2020, Barzanjeh2019, Barzanjeh2022, Arnold2020} is served to generate entanglement between microwave photons (signal mode) and optical photons (idler mode). The resulting optical mode is carefully preserved at the receiver, intended for subsequent joint measurements. Simultaneously, the microwave mode is utilized for probing the target region. The reflected microwave signal passes through a second EOM cavity specially designed to convert and phase conjugate the microwave signal to optical photons, see Fig. \ref{Microwaveqi}b. Subsequently, a joint measurement is executed on the converted signal and the retained idler beams (now both in the optical domain), aligning perfectly with the fundamental principles delineated in Gaussian quantum illumination theory.

\section{Experimental Realization of Quantum Illumination: Microwave and optical domains}

 \subsection{Microwave quantum radar based on Homodyne and Heterodyne measurements}
In recent years, a series of experiments have been conducted to examine the principles of microwave quantum illumination under highly controlled laboratory conditions and using linear type detection (Homodyne and Heterodyne) combined with heavy postprocessing. A quantum radar prototype described in \cite{Chang et al. 2019, Luong et al. 2019} experimentally validated the concept of quantum-enhanced noise radar, also referred to as "quantum two-mode squeezing radar" in \cite{Luong et al. 2019}. This experimental demonstration shows an improved performance achieved by utilizing quantum-entangled signals and idler beams, especially when compared to a classical counterpart employing a two-mode noise radar system operating under identical conditions. Similarly, in the experiment led by Barzanjeh and colleagues \cite{Barzanjeh et al.2019}, a systematic comparison was carried out between quantum illumination, coherent state illumination, and classical noise with classical correlations. This comparison ensured that all illumination sources were carefully prepared with identical specifications and under matching conditions. The primary objective of this experiment was to assess the performance of quantum illumination when faced with a low-reflectivity target positioned in a space-like environment. 

Both experimental protocols roughly involve the following sequential steps:
  \begin{figure}[t]
\begin{center}
 \includegraphics [width=1\linewidth]{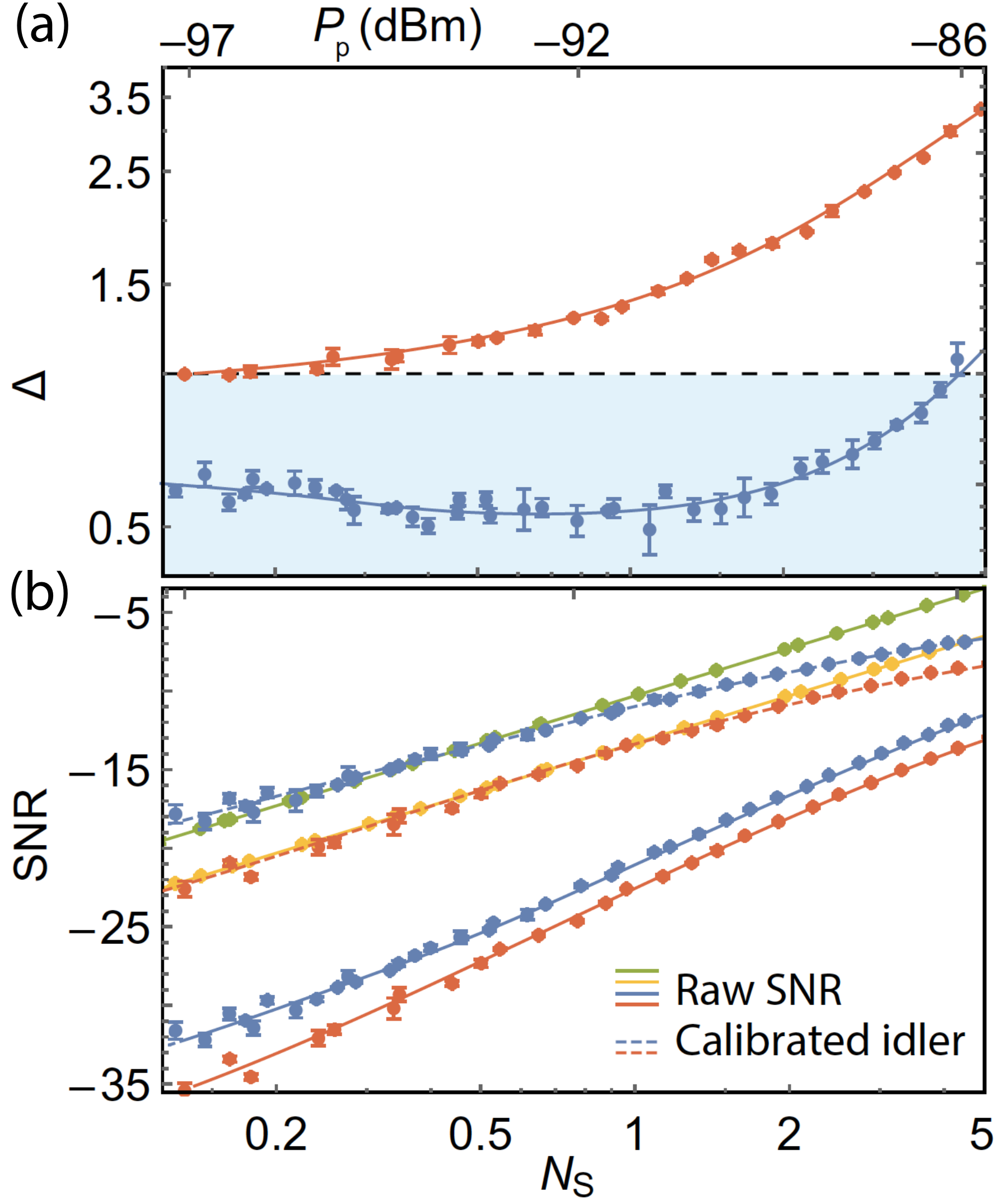}
\caption{Entanglement and Quantum Illumination (QI):(A) The graph displays the measured entanglement parameter $\Delta$ for the JPC output (in blue) and classically correlated noise (in orange). It illustrates how $\Delta$ varies with the inferred signal photon number $N_S$ at the JPC output and the input pump power $P_p$ to the JPC. (B) This plot compares the measured single-mode signal-to-noise ratio (SNR) of Quantum Illumination (QI, solid blue) with symmetric classically correlated illumination (CI, solid orange), coherent-state illumination with homodyne detection (solid green), and heterodyne detection (solid yellow). Additionally, it shows the inferred SNR of calibrated QI (dashed blue) and CI (dashed orange) concerning the signal photon number $N_S$. These measurements are made assuming a perfectly reflective object and a 5 $\mu$s measurement time. Data points, both measured and inferred, are represented by dots, while solid and dashed lines represent theoretical predictions \cite{Barzanjeh et al.2019}.}
\label{SNRBarzanjeh}
\end{center}
\end{figure}

\begin{enumerate}
\item Two entangled microwave beams are generated directly from a JPA. The entanglement between these modes has been verified by inferring Duan-Zoller inequality, see Fig. \ref{SNRBarzanjeh}a. 

\item The idler beam is amplified and measured using heterodyne detection.

\item In parallel, after amplification, the signal beam is sent to probe the region where the target is located.

\item Classical digital filtering techniques are used in the detection and storage of the received beam.

\item 
The detection process involves measuring the linear quadratures and employing filtering to compare the two signals. This procedure enables the implementation of a phase-conjugate receiver digitally, effectively utilizing the correlations between the idler and signal.
\end{enumerate}

For the classically correlated light illumination, the process is analogous, maintaining the same conditions of temperature, energy, and power for the idler/signal generation as in the quantum illumination case. 

 In Barzanjeh's experiment, the target remained fixed at a distance of up to $1$ m from the transmitting antenna. Both the antennas and the target were maintained at room temperature. Heterodyne detection was employed to capture the reflected signal, and subsequently, the two measurements underwent post-processing to calculate the signal-to-noise ratio.

In both experiments, the generation and amplification of entanglement occurred under cryogenic conditions at 7 millikelvins. The outcomes of Barzanjeh et al. experiment showed an improvement in signal-to-noise ratio when employing quantum-entangled microwave radar, in contrast to symmetric and suboptimal classical and coherent illumination schemes \cite{Barzanjeh et al.2019}:
 \begin{itemize}
 \item Enhancement up to $3$ dB in the SNR in the low-intensity regime with respect to sub-optimal, symmetric classical illumination, see Fig. \ref{SNRBarzanjeh}b.

 \item Under the simulated situation of perfect idler photonumber detection (inferred in postprocessing), the SNR for quantum illumination is up to $4$ dB higher than classical illumination, defeating also coherent light illumination with heterodyne detection.

 \item  For signals with photon number $N_S >4.5$, there is no advantage from entangled radiation compared to coherent light illumination.

 \item
 For $N_S<0.4$, the SNR of quantum illumination is 1 dB larger than the coherent state using homodyne detection, see Fig. \ref{SNRBarzanjeh}b. 
 \end{itemize}

Both quantum radar experiments make use of entangled radiation for their illumination processes. However, the mechanisms for receiving the signal and idler modes rely on traditional digital techniques and classical filter methods. The conclusion drawn from these experiments is that while such hybrid protocol designs may offer advantages in terms of integration time, they often involve a trade-off that could potentially result in a loss of SNR enhancement due to the use of linear voltage (quadrature) measurements. Conversely, employing digital detectors allows for the storage of idler mode and can potentially eliminate the need for quantum memory. As explained above, Barzanjeh's experiment simulates the situation involving ideal idler photon number detection, a requirement in quantum illumination, resulting in a quantum advantage when compared to the equivalent classical (coherent state) benchmark.

 \subsection{Microwave quantum radar based on photocounting}
Microwave quantum radar can also be realized through direct photon counting, as demonstrated in \cite{Assouly et al. 2023}. This approach resulted in a performance improvement, exceeding $20\%$ compared to traditional radar systems. The experimental methodology described in this work focused on the concurrent measurement of entangled signal and idler microwave photon states immediately after the signal's reflection from the target. To address the issue of idler loss while preserving the quantum advantage, the idler mode has been stored in a high-quality superconducting resonator.  

The signal mode promptly exited its resonator, sent to the target, and upon interacting with the target, returned with an attenuation. The target configuration comprised a circulator and a flux-tunable notch filter, followed by a $12$ m-long coaxial cable. This setup allowed real-time adjustment of the target's reflectivity. Following this phase, the reflected signal was combined with thermal noise introduced through a weakly coupled auxiliary line. This noise was generated at room temperature by amplifying the Johnson–Nyquist noise of a $50\,\Omega$ resistor with tunable gain. Employing parametric amplification and direct photodetection, as suggested by Guha and Erkmen \cite{Guha Erkmen 2009}, the system's error exponent was inferred.

\subsection{The current limitations of microwave quantum radar}
The pioneering experiments in microwave quantum radar, as discussed previously, have undergone thorough examination, particularly in terms of their tangible advantages over classical radar protocols \cite{Shapiro2019}. This scrutiny has led to the identification of key challenges hindering the practical application of quantum illumination in quantum radar, particularly within the microwave frequency range.  These challenges encompass issues such as idler storage, bandwidth constraints, strict energy prerequisites for demonstrating benefits, and the imperative need for cryogenic environments in microwave quantum radar.

In the context of microwave quantum illumination, the bandwidth limitation arises primarily from the conventional method of microwave entanglement generation, i.e., through JPA technology. This limitation severely restricts the maximum achievable range for quantum radar systems. Consequently, expanding the bandwidth has emerged as a pivotal objective in microwave quantum illumination. Recent discussions have explored alternative approaches for generating entanglement with broader bandwidths, as outlined in \cite{Livreri et al. 2022}. This protocol leverages the use of Josephson Traveling Wave Parametric Amplifiers (JTWPA) for entanglement generation. Unlike the JPA, the JTWPA facilitates the generation of significantly wider bandwidth signals, particularly at X-band radar frequencies. For instance, at a pump frequency of 12 GHz, the bandwidth can extend to 3-4 GHz. Ref \cite{Livreri et al. 2022} also addresses the cryogenic requirements necessary to circumvent noise during generation, albeit still necessitating temperatures in the range of approximately $10$ mK. Recent works have explored the utilization of JTWPA, maintained at millikelvin temperatures, to generate entangled photon pairs with substantial bandwidth, as explained in \cite{Esposito2022} and \cite{Qiu et al. 2023}. 

\section{Quantum LiDARS based on the protocol of quantum illumination and beyond}
When quantum illumination is applied in optical frequency ranges for target detection, it leads to the development of quantum sensing protocols in the domain of quantum LiDARs. Implementing quantum illumination for LiDAR applications in optical frequencies offers significant advantages compared to microwave quantum radar. It eliminates the requirements for cryogenic cooling systems, which are typically necessary for microwave quantum systems. Additionally, It can provide superior resolution and enhanced accuracy when compared to radar, particularly when it comes to detecting small objects and performing effectively under adverse weather conditions.

 \begin{figure}[t]
 \begin{center}
 \includegraphics [width=1\linewidth]{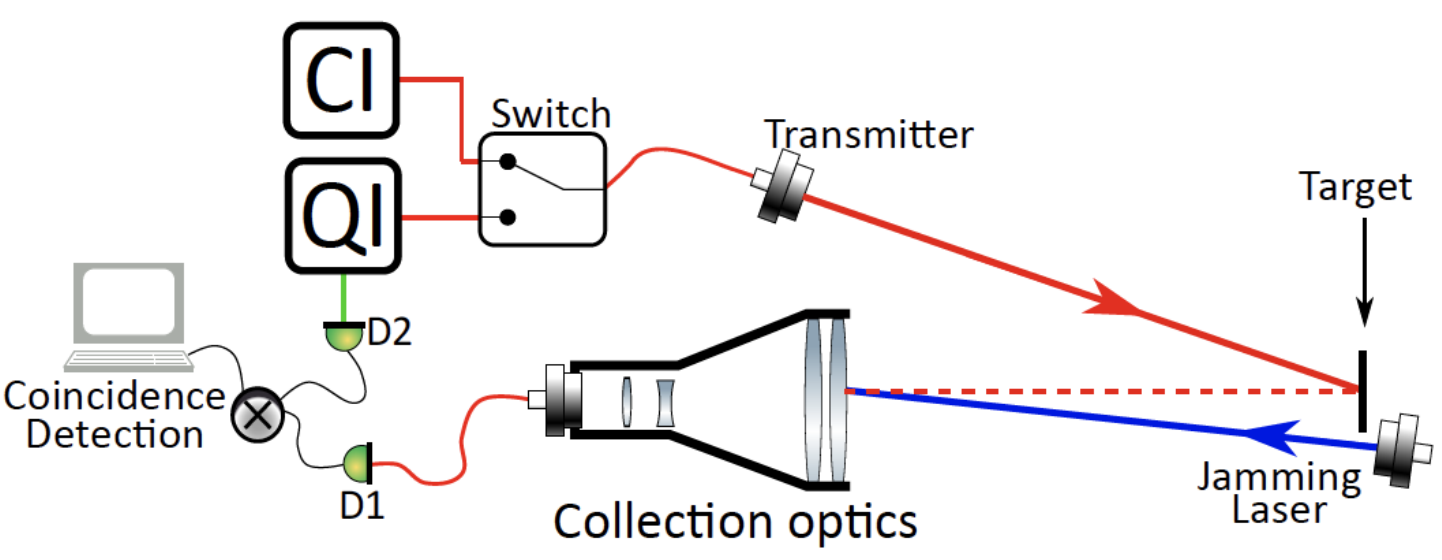}
\caption{An experimental diagram depicting the setup used in Ref. \cite{England Balaji Sussman 20019}: A light source, which can emit either quantum (QI) or classical (CI) light, is directed toward the transmitter, which then illuminates a target. The light scattered from this target is gathered by the collection optics and guided toward detector D1. In the case of quantum illumination (QI), the herald beam is detected by detector D2, and measurements involving coincidences between the two photons are conducted. Additionally, a second laser, referred to as the jamming laser, serves as a background light source.}
\label{Lidar00}
\end{center}
\end{figure}

\subsection{Quantum LiDAR}
 The initial demonstrations of quantum illumination were indeed accomplished in the optical domain, where the presence of highly efficient direct photon detection methods allows for the practical realization of quantum illumination experiments with maximum potential benefits. Two notable examples of quantum LiDAR are the works by Lopaeva \cite{Lopaeva et al.} and England \cite{England Balaji Sussman 20019}. These experiments entailed a performance evaluation of standard detector systems utilizing classical light in comparison to systems employing quantum illumination. 

The experimental methodology employed by England closely aligns with that of Lopaeva. In this case, the Signal-to-Noise Ratio was determined using the following expression
\begin{align}
\text{SNR}=\,\frac{N_{in}-N_{out}}{N_{out}}
\end{align}
This expression applies to both classical and quantum illumination systems. Here, $N_{in/out}$ represents the count of detected photons when the target is present/absent.

Both experiments employed single-photon detection for classical illumination and coincidence detection for quantum illumination. In both studies, the advantage of quantum illumination over classical thermal light illumination was demonstrated, and this advantage was evident in scenarios characterized by strong external interference (induced by jamming signal) as well as in low-noise environments.

Despite these similarities, significant distinctions exist between the experiments. Lopaeva employed a perfectly reflective target, while England used a diffusive target, demonstrating a higher level of realism for the application of quantum LiDAR in practical situations, see Fig. \ref{Lidar00}. 

Beyond the range issue discussed earlier, which affects quantum radar based on quantum illumination across all frequency ranges, a main concern hindering the application of Gaussian quantum illumination in realistic LiDAR scenarios is related to the condition of bright background noise, where $N_B \gg 1$ is assumed. In the optical regime, where sources of entangled beams via parametric down-conversion methods are relatively accessible, this condition is not met under normal skylight conditions, where $N_B(optical\, sky) \ll 1$. However, in situations where jamming is a significant concern, quantum illumination can offer sensitivity advantages, as demonstrated experimentally in \cite{Zhao et al. 2022}. This work shows that quantum LiDAR can provide a better SNR even in the presence of jamming systems.

Recent progress in quantum LiDAR, particularly in the context of quantum illumination protocols, presents significant advancements with implications for metrological applications, as we will discuss further. Regarding sensitivity enhancement, we can highlight the work by Frick et al. \cite{Frick McMillan Rarity 2020}. They have introduced a quantum LiDAR protocol that stands out for its superior range-finding capabilities and its unique ability to provide camouflage against competing thermal sources, capitalizing on an additional aspect of quantum illumination.

\subsection{Quantum LiDAR and phase stability}
The stability of phase plays a central role in both QI and optimal classical coherent detection. However, practical constraints often make it exceptionally challenging to maintain phase coherence among interacting beams. Consequently, in practical classical LiDAR applications, intensity-based detection schemes are generally preferred due to their convenience. As a result, the use of optimal coherent states as a benchmark is less common and may not provide a representative comparison.

To address this challenge in the optical domain, alternative quantum-enhanced target detection methods that are insensitive to phase variations have been proposed and investigated. These methods leverage quantum correlations in both intensity and time domains \cite{Liu2019}. The work of England et al. \cite{England Balaji Sussman 20019} can also be categorized into this type of quantum LiDAR. Phase-insensitive detection approaches offer a simplified implementation process, making them attractive for scenarios where significant phase noise is introduced during operation. References \cite{Liu2020, He2020} have explored phase-insensitive detection approaches, demonstrating their exceptional noise resilience compared to classical phase-insensitive methods. It's important to note, however, that these methods have exhibited enhanced performance over classical phase-insensitive counterparts within specific ranges of noise power levels.

 \begin{figure}[t]
\begin{center}
 \includegraphics [width=1\linewidth]{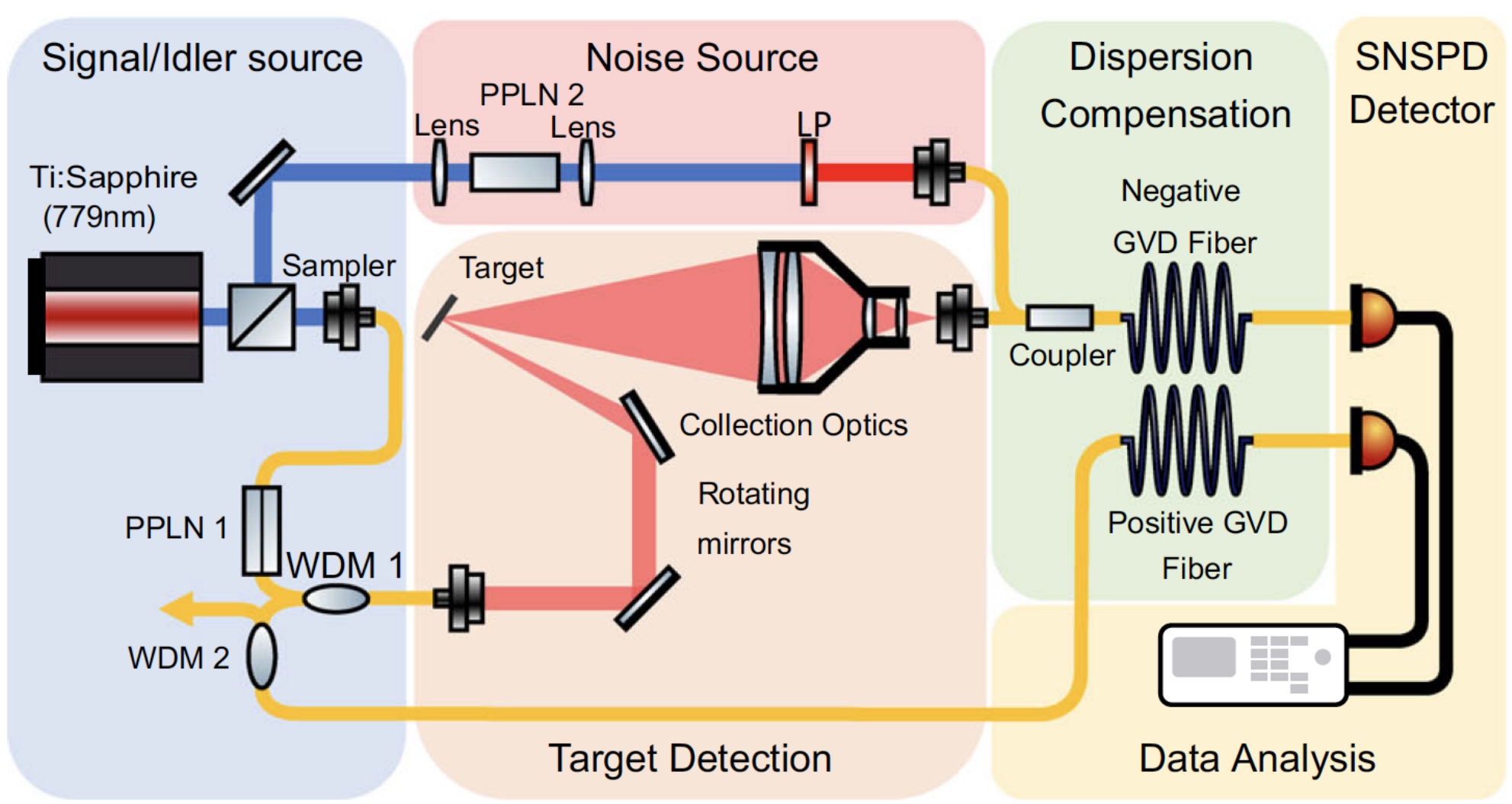}
\caption{Schematic of the dispersed non-classical target detection protocol used in Ref. \cite{Blakey}. Here: Periodically Poled Lithium Niobate (PPLN), Group Velocity Dispersion (GVD), and Superconducting Nanowire Single-Photon Detector (SNSPD).}
\label{Lidar}
\end{center}
\end{figure}

The utilization of quantum temporal correlations proves advantageous in LiDAR applications \cite{Blakey}. By conducting measurements within a rotated basis that encompasses both time and frequency domains, this technique facilitates the amplification of temporal uncertainty between the signal and idler, all while preserving their correlation. Consequently, these correlations can be harnessed to their full potential, greatly enhancing the ability to discriminate targets from background noise. The application of an appropriate temporal window serves to effectively filter out noise components that no longer overlap with the signal. As a result, this approach yields a significant improvement of up to $43$ dB in the signal-to-noise ratio when compared to a phase-insensitive classical target detection method employing an equivalent probe power level.

 \begin{figure}[t]
\begin{center}
 \includegraphics [width=1\linewidth]{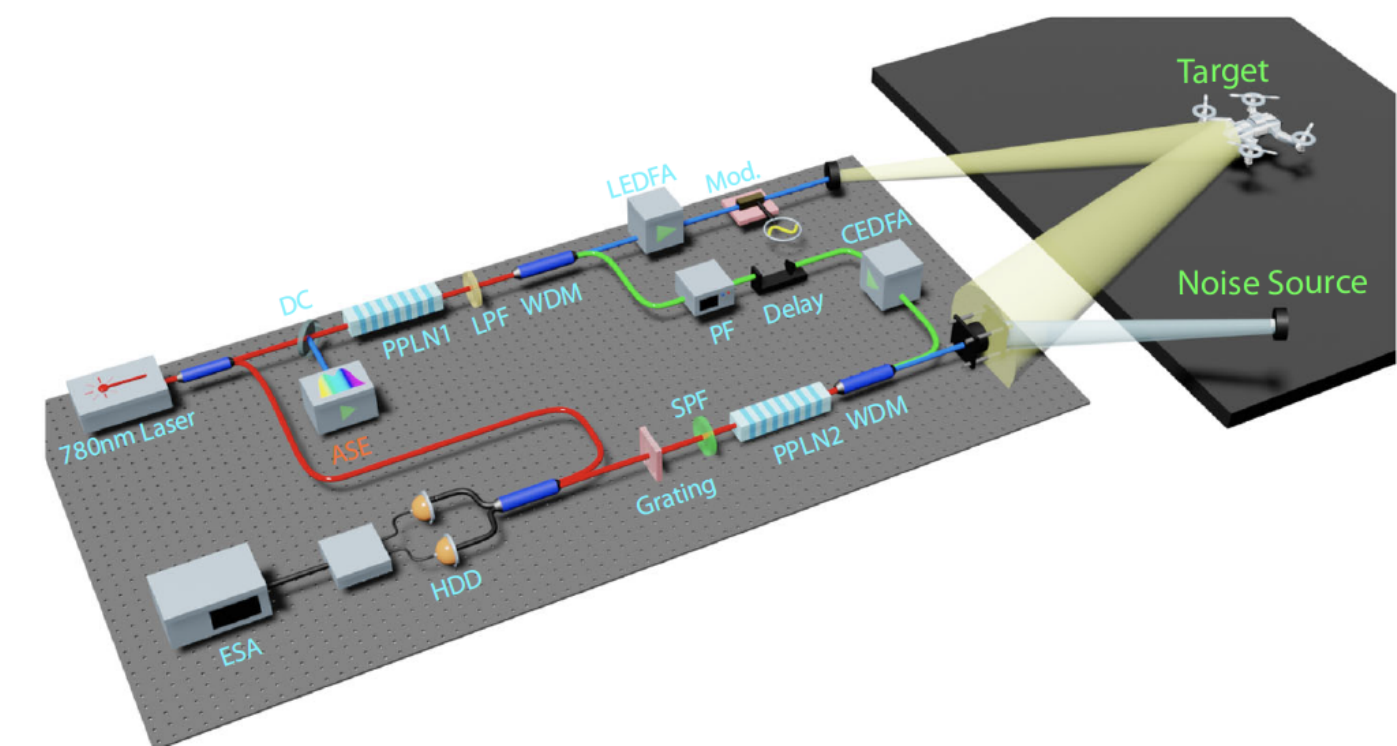}
\caption{The experimental setup of the chaotic-quantum frequency conversion LiDAR system used in Ref. \cite{Liu}.
DC dichroic combiner, WDM wavelength demultiplexer, Mod phase modulation for frequency shifting, PF programmable amplitude-phase filter, HDD balanced homodyne detection, ESA electric spectrum analyzer, SPF/LPF short/long pass filter, ASE amplified spontaneous emission, PPLN1, 2 periodically poled lithium niobate waveguide for DFG and SFG, Target: quadcopter drone on a separate table, CEDFA, LEDFA: C and L band erbium-doped fiber amplifiers.}
\label{ClassicLidar}
\end{center}
\end{figure}

\subsection{Quantum-inspired LiDAR}
While quantum illumination outperforms classical LiDAR with equivalent signal power, its practicality is limited due to challenges in increasing the flux of entangled light sources. This makes it challenging for QI to meet the demands of real-world sensing, requiring high probe power for extended detection ranges. One possible avenue to resolve this issue is to develop quantum-inspired classical LiDARs. Reference \cite{Liu} presents a LiDAR design, featuring a similar setup and operational principle as quantum illumination, see Fig. \ref{ClassicLidar}. However, it employs a classical time-frequency correlation source with substantial power, greatly enhancing its potential for achieving extended operating ranges. This research demonstrates that through classical time-frequency correlation, probe light characterized by random and chaotic time-frequency attributes can be selectively converted into a single frequency with nearly perfect quantum efficiency. Simultaneously, in-band noise with identical time-frequency characteristics can be effectively reduced to negligible levels. Note that this noise reduction technique bears conceptual similarity to quantum frequency conversion (QFC), a quantum information processing method where a specific time-frequency mode (corresponding to LiDAR probe light) is selectively isolated from the rest of the spectrum (representing indistinguishable noise) while preserving its quantum properties.

\section{Extensions of Quantum Illumination protocol} 
\begin{figure}[h]
\begin{center}
 \includegraphics [width=1\linewidth]{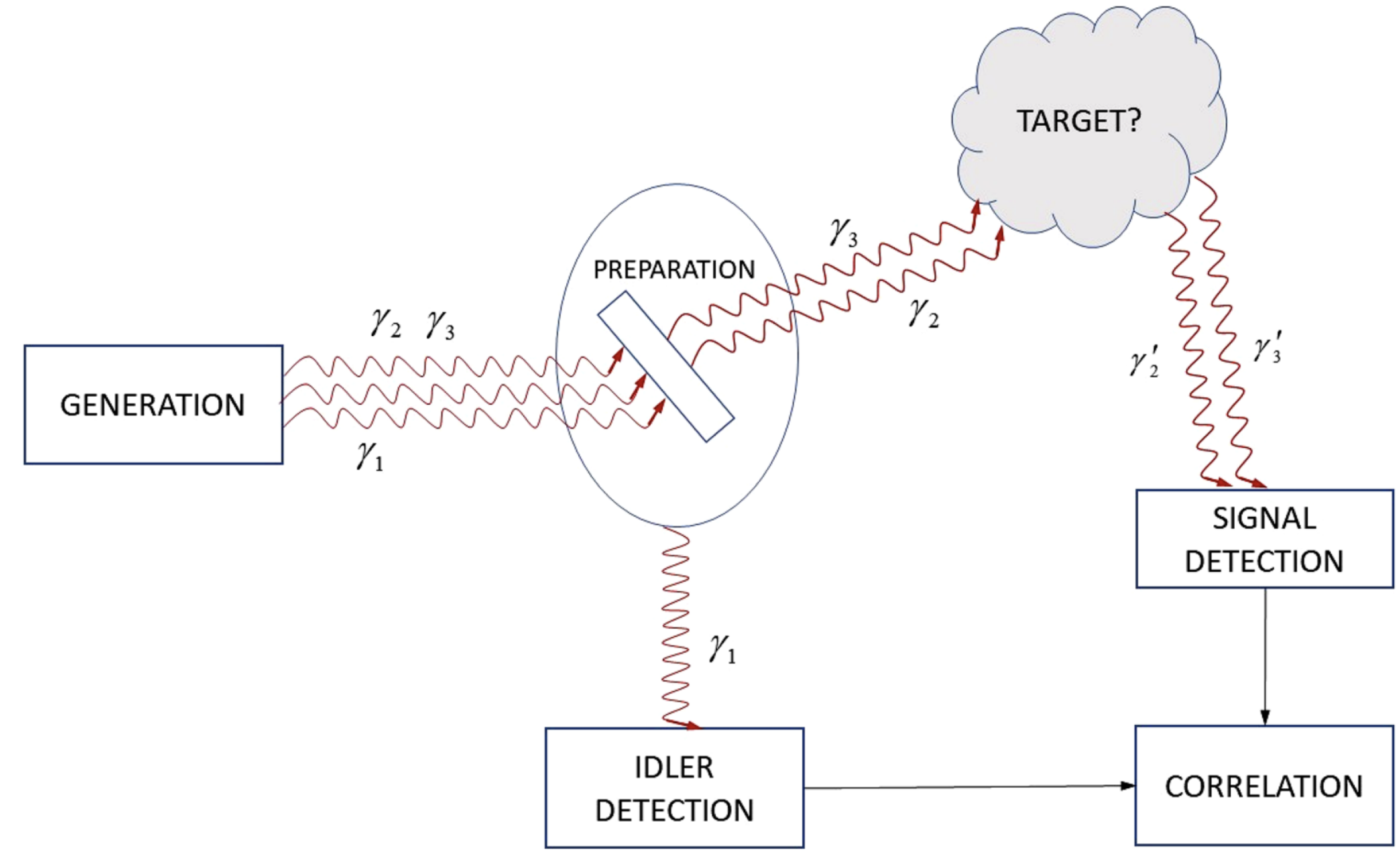}
\caption{ Quantum illumination with three entangled photon states  \cite{Ricardo2020b, Ricardo 2023}.}
\label{directmeasurement}
\end{center}
\end{figure}

A major challenge in employing quantum illumination for radar applications pertains to the exceedingly low signal-idler mode (beam) intensity required to gain an advantage over coherent light illumination. Partially motivated by the necessity for a weak signal-idler system, there has been an exploration of extending the theory to encompass multiple entangled photons, as discussed in \cite{Ricardo2020b, Ricardo 2023}. The core concept involves broadening the quantum illumination protocol to contain scenarios where the signal beam comprises multiple photons entangled with the idler photon. This protocol circumvents the inherent sensitivity limitations of two-mode quantum illumination \cite{de Palma Borregaard, Nair Gu 2020, Su-Yong Lee et al. 2022}, exhibiting improved sensitivity compared to two-mode Gaussian quantum illumination (see Fig. \ref{directmeasurement}). It's important to note that the advantage over two-mode Gaussian quantum illumination could depend on the specific characteristics of the quantum states employed, as elaborated upon in discussions in \cite{Su-Yong Lee et al. 2021, Noh Lee Lee}. 
Another perspective on multi-entangled photon states in quantum illumination has been presented in \cite{Jung and Park 2022}. In this protocol, the signal-idler system is characterized by three-mode entangled Gaussian states, with the signal state composed of one mode while the idler states are two-mode photon states. The comparative analysis of the probability of error shows that for $N_S < 0.295$, the error probability is lower than that observed in two-mode Gaussian quantum illumination.

The protocol of quantum illumination has also been extended beyond Gaussian state \cite{Daum et al. 2o21}. Instead of relying on the Schrödinger equation, the authors opted for the Belavkin-Zakai equation to describe the dynamics. Additionally, addressing the challenge of increasing signal intensity, Djordjevic has outlined potential solutions in two related papers \cite{Djordjevic 2022a, Djordjevic 2022b}. These protocols involve an increase in the parameter $N_S$ compared to Gaussian quantum illumination, with the advantage of the proposed scheme becoming apparent at moderate Signal-to-Noise Ratios, where $\text{SNR} = N_S / (2N_B + 1)$. 

The fundamental protocol of quantum illumination traditionally demands the use of a maximally entangled Gaussian state. However, recent research \cite{Karsa et al. 2020} has embarked on relaxing the stringent transmitter requirements in quantum illumination. This exploration involves shifting away from the conventional maximally entangled two-mode squeezed vacuum source toward a more versatile quantum-correlated Gaussian source, which may even exhibit separability. The outcomes of these investigations reveal that quantum advantages can still be harnessed when employing Gaussian sources that do not necessarily possess maximum entanglement. These advantages are particularly pronounced in scenarios characterized by short detection ranges, where the adverse effects of spherical beam spreading result in minimal signal loss. This development holds significant promise, particularly in the context of quantum radar applications, where mitigating loss due to beam spreading is a critical factor for success.

\section{Quantum radar and LiDAR: accuracy enhancing protocols}
Numerous protocols have been devised to harness quantum entanglement for achieving the utmost precision in parameter determination, and these foundational principles have been solidified in the general theory to reach standard quantum limit \cite{GiovannettiLloydMaccone}. In this section, we describe several approaches that leverage these concepts for radar and LiDAR applications. Consequently, one of the secondary advantages that quantum radar protocols can offer over classical radar lies in parameter estimation and metrological applications \cite{PhysRevLett.106.090504}. This aspect was initially recognized during the development of quantum interferometric radar, as detailed in reviews such as \cite{Marco Lanzagorta 2011, Gallego Torrome Bekhtil Knott 2021}. Nonetheless, we will also explore other protocols that facilitate enhancements in this context.

\subsection{Range determination accuracy enhancement }

\begin{figure}[t]
\begin{center}
 \includegraphics [width=1\linewidth]{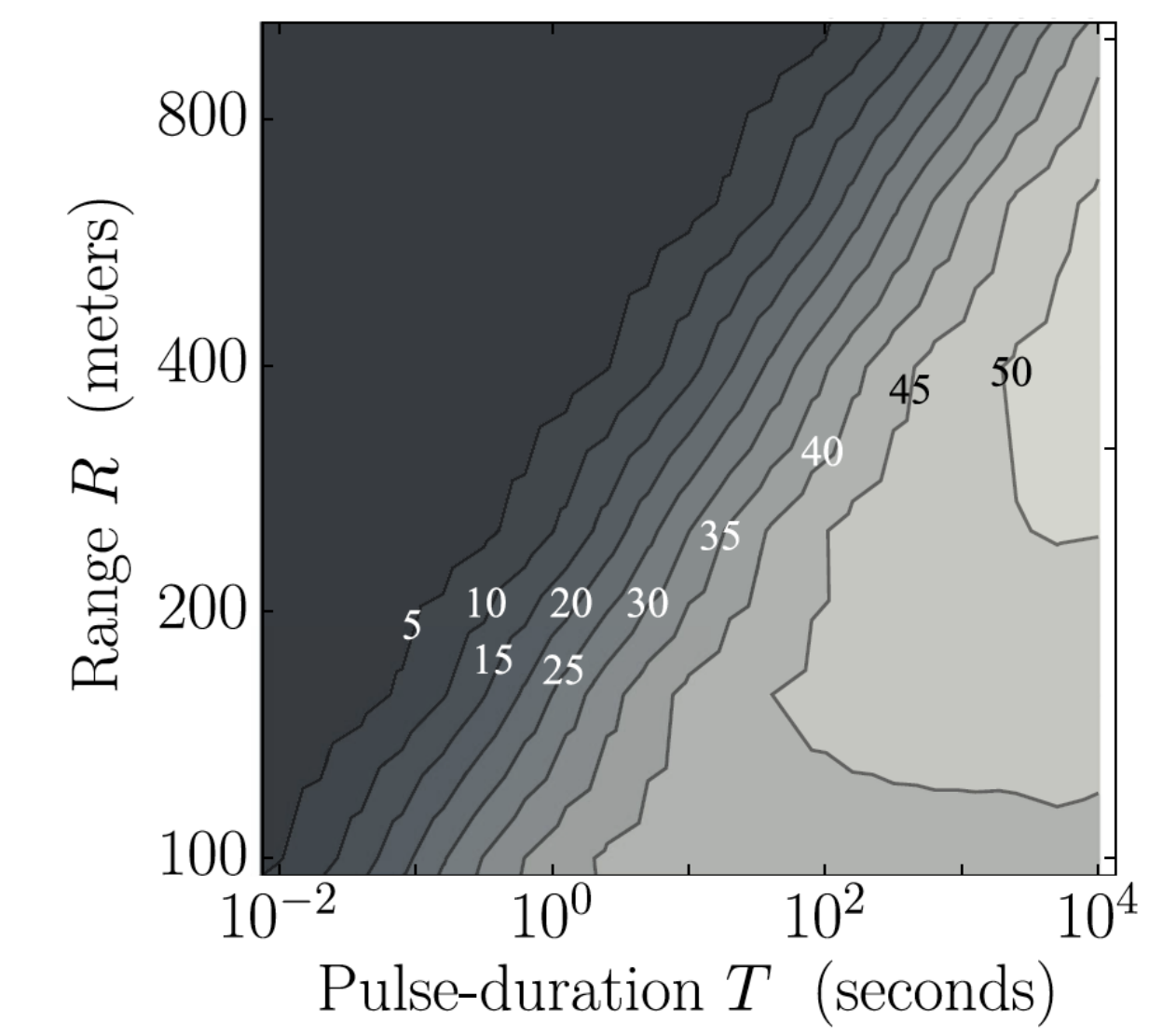}
\caption{Enhancement in accuracy in range determination using quantum illumination. Mean-squared range-delay accuracy
advantage (in dB) \cite{Zhuang Shapiro 2022}.}
\label{fig:RangeIntegrationtime}
\end{center}
\end{figure}

A recent advancement in the domain of quantum illumination involves the revelation of enhanced accuracy in mean square range-delay determination through the utilization of entangled states, as demonstrated in \cite{Zhuang Shapiro 2022}. The authors devised a continuous-time approach to address target ranging in quantum illumination and compared its performance to that of a classical pulse-compression radar. Both systems rely on time-of-flight measurements to estimate target range, leading to SNR thresholds below which their range-delay measurement accuracy is worse than their Cramér-Rao bound limit. In this context, the quantum radar exhibits a lower threshold SNR by 6 dB compared to the classical radar. Consequently, when the quantum radar operates at its threshold SNR, its mean-squared range-delay accuracy can surpass that of its classical counterpart by tens of dB, see Fig. \ref{fig:RangeIntegrationtime}. The main difficulty in the implementation of this protocol in practical settings resides in the fact it requires a large integration time to perform the measurement.

Several other research works have identified instances where quantum illumination offers advantages in parameter determination. For instance, in Ref \cite{Reichert et al. 2022} quantum illumination is employed to ascertain the velocity of a non-cooperative target using the Doppler effect, exhibiting improved accuracy in velocity determination. This protocol is applicable across both optical and microwave frequencies. It is worth noting that quantum mechanics imposes fundamental limits on the estimation of both range and velocity, as explained in \cite{GiovannettiLloydMaccone, Huang et al. 2021}. These findings on quantum illumination suggest a connection between the enhancement of quantum sensitivity and the enhancement of accuracy in parameter determination \cite{Zhong et al. 2023}.

\subsection{Enhancement in accuracy in quantum LiDAR}
At the LiDAR regime, significant progress has been made in the field of metrology and target parameter determination, which warrants discussion. From a theoretical perspective, the work by Huang et al. \cite{Huang et al. 2021} introduces a protocol for determining both the range and velocity of targets utilizing the Doppler effect. In this protocol, the signal model for single-photon illumination consists of a Gaussian state packet with the minimum duration-spread product, while for two-photon entangled states, it employs a Gaussian state akin to those generated in SPDC, featuring frequency-entangled photons. The methodology facilitates range and velocity determination by evaluating the round-trip time of the signal. Importantly, this theory can be extended to address scenarios involving multiple concurrent targets. A closely related approach has been pursued in \cite{Reichert et al. 2022}, demonstrating that two-mode Gaussian states offer an advantage in Doppler frequency determination \cite{Zhao et al. 2022}. 


\subsection{Quantum interferometric radar: principles, recent experiments and challenges}

Interferometric quantum radar is built upon the formal analogy between radar detection procedures and Mach-Zehnder interferometry. 
In a Mach-Zehnder interferometer configuration utilizing coherent light, an initial light beam is split into two separate paths by a beam splitter. Each of these beams follows separate optical pathways, and upon their reunion, an interference measurement is conducted. The difference in the lengths of these optical paths introduces a phase discrepancy $\varphi$ between the states upon their reunion.  By measuring the intensity of the beams at the receiver, which involves observing the statistical distribution of individual photons arriving at the detector or employing alternative detection techniques, we can acquire information about $\varphi$. Analyzing the phase information allows for the retrieval of target parameters.
When coherent light composed of $N$ independent photons is employed as the source beam, the error in estimating $\varphi$ decreases asymptotically with the number of photons $N$ in the beam as $1/\sqrt{N}$. This outcome directly stems from the statistical independence of the $N$ individual photons comprising the light beam.

 Assuming the source beam is coherent light, it is divided into two beams. One beam is directed along an optical path to serve as the signal beam, while the second beam, known as the idler beam, is retained in the detector throughout the experiment. After the signal beam returns, it is combined with the idler beam, and an interferometric experiment, involving a joint measurement, is conducted. Similar to Mach-Zehnder interferometry, the phase information $\phi$ can be extracted by measuring the intensities of detected photon distributions. Quantum metrology theory \cite{GiovannettiLloydMaccone2001, GiovannettiLloydMaccone} establishes that the error in estimating $\varphi$ translates to an error in estimating the target range, following the form:

\begin{align}
\delta R=\,\mathcal{O}\left(\frac{1}{\Delta \omega \sqrt{N}}\right),
\label{optical path sensitivity}
\end{align}
where $\Delta \omega$ is the signal bandwidth. An interferometric radar works as a Mach-Zehnder interferometer, but where one of the arms of the interferometer (corresponding to the signal beam) is much larger than the other (see \cite{Marco Lanzagorta 2011}, section 5.2).

The analogy between quantum interferometry and interferometric radar can be extended to the case when beams are composed of entangled quantum states of light. NOON-states are among the types of quantum entangled states that have been considered for quantum sensing applications \cite{Boto et al.,Marco Lanzagorta 2011}. A NOON-quantum state has this form:
\begin{align*}
|\psi\rangle & =\,\frac{1}{\sqrt{2}}\,\left(\frac{(a^\dag_1)^N}{\sqrt{N !}}\otimes I_{d_2}+\,I_{d_1}\otimes \frac{(a^\dag_2)^N}{\sqrt{N !}}\right)|0\rangle_1|0\rangle_2\\
& \equiv  \,\frac{1}{\sqrt{2}}\left(|N0\rangle+\,|0N\rangle\right),
\end{align*}
where $|0\rangle_1|0\rangle_2$ is the vacuum state and $\{a_i,\,i=1,2\}$ are the annihilation operators for the photons pass by arm $1$ or arm $2$. The effect of a phase shift $\varphi$ along one of the arms implies that the quantum state is of the form
\begin{align}
|\psi\rangle \equiv  \,\frac{1}{\sqrt{2}}\left(|N0\rangle+\,e^{i N\varphi}\,|0N\rangle\right)
\label{NOONstates}
\end{align}
It can be demonstrated that the error in estimating the phase $\varphi$ when employing a specific observable for the NOON-state is approximately on the order of $\delta \varphi \sim 1/N$ \cite{GilbertWeinstein2008,Marco Lanzagorta 2011}. This implies a corresponding level of precision in estimating the target's range $R$ which differs significantly from the asymptotic precision of approximately $1/\sqrt{N}$ when utilizing coherent states. This precision at the level of $1/N$ in range estimation is commonly referred to as the  {\it Heisenberg limit} \cite{MargulisLevitin}.

The theory explained above is formulated under ideal circumstances. Nonetheless, it's important to note that achieving the Heisenberg limit using NOON-states becomes unattainable in the presence of attenuation caused by absorption and scattering of the beam within the propagation medium \cite{Marco Lanzagorta 2011, GilbertHamrick, GilbertWeinstein2008, Gilbert Hamrick Weinstein2008}
\begin{align}
\nonumber |\psi\rangle  \equiv \, &\frac{1}{\sqrt{2\,N!}}\,e^{-\left(i \, \eta_1\,\frac{\omega}{c}+\,\kappa_1 \omega/2\right)\,\left(N\,L_1\right)}\,\left(\hat{a}^\dag_1\right)^N\,|0\rangle_1|0\rangle_2\\
&  +\frac{1}{\sqrt{2\,N!}}\,e^{-\left(i\,\eta_2\, \frac{\omega}{c} +\,\kappa_2\omega/2\right)\,\left(N\,L_2\right)}\,\left(\hat{a}^\dag_2\right)^N\,|0\rangle_1|0\rangle_2,
\label{attenuate NOON state model}
\end{align}
where $\eta_i$ are refractive indices of the paths $i=1,2$, $\kappa_i(\omega)$ are attenuation indices, $\omega$ is the frequency of the radiation and $c$ is the speed of light in vacuum. For $L_2\approx\,L_1$ and $\kappa_1 \ll\kappa_2$, the model implies an exponential attenuation along the path $2$ with respect to the path $1$,
\begin{align*}
|\psi\rangle \to\,\frac{1}{\sqrt{N!}}\,e^{-\left(i \eta_1\,\frac{\omega}{c}+\,\kappa_1(\omega)/2)\right)\,\left(N\,L_1\right)}\,\left(\hat{a}^\dag_1\right)^N\,|0\rangle_1|0\rangle_2+ \delta |\psi\rangle,
\end{align*}
where $\delta |\psi\rangle$ is an exponentially attenuated state with respect to the first term.


Quantum interferometric radar could also be realized by means of  entangled states as described in \cite{Yurke et al. 1986, Dowling, GiovannettiLloydMaccone},
\begin{align}
|\Psi\rangle =\,\frac{1}{2}\Big(|N_+\rangle_A\,|N_-\rangle_B+\,|N_-\rangle_A\,|N_+\rangle_B\Big)
\label{N+N- states}
\end{align}
where $ N_{\pm}=\,\frac{N\pm 1}{2}$. Similar to NOON-states, the measurement error of $\varphi$ scales proportionally with $1/N$ demonstrating a reduction in the estimation error by a factor of $\sqrt{N}$ compared to coherent light.

\subsection{Quantum radar of Maccone and Ren: principle, recent developments and challenges}
A radar system aims not only to detect a target but to provide the parameters of the target as range, speed, and size. While quantum illumination has been explored as a quantum radar protocol, it has been made manifest that the usual quantum illumination protocol cannot work as a radar protocol, because it assumes previous knowledge of the target range.

In Maccone-Ren's protocol, quantum states with $N$ entangled photons are prepared. For each individual state, all the $N$ photons are sent to explore a region of spacetime possibly containing a non-cooperative point-like target. Therefore, the difference with quantum illumination protocols is that all the entangled photons are sent to explore the target and none is preserved as an idler.
\begin{figure}[t]
\graphicspath{{Images/}}
\begin{center}
 \includegraphics [width=1\linewidth]{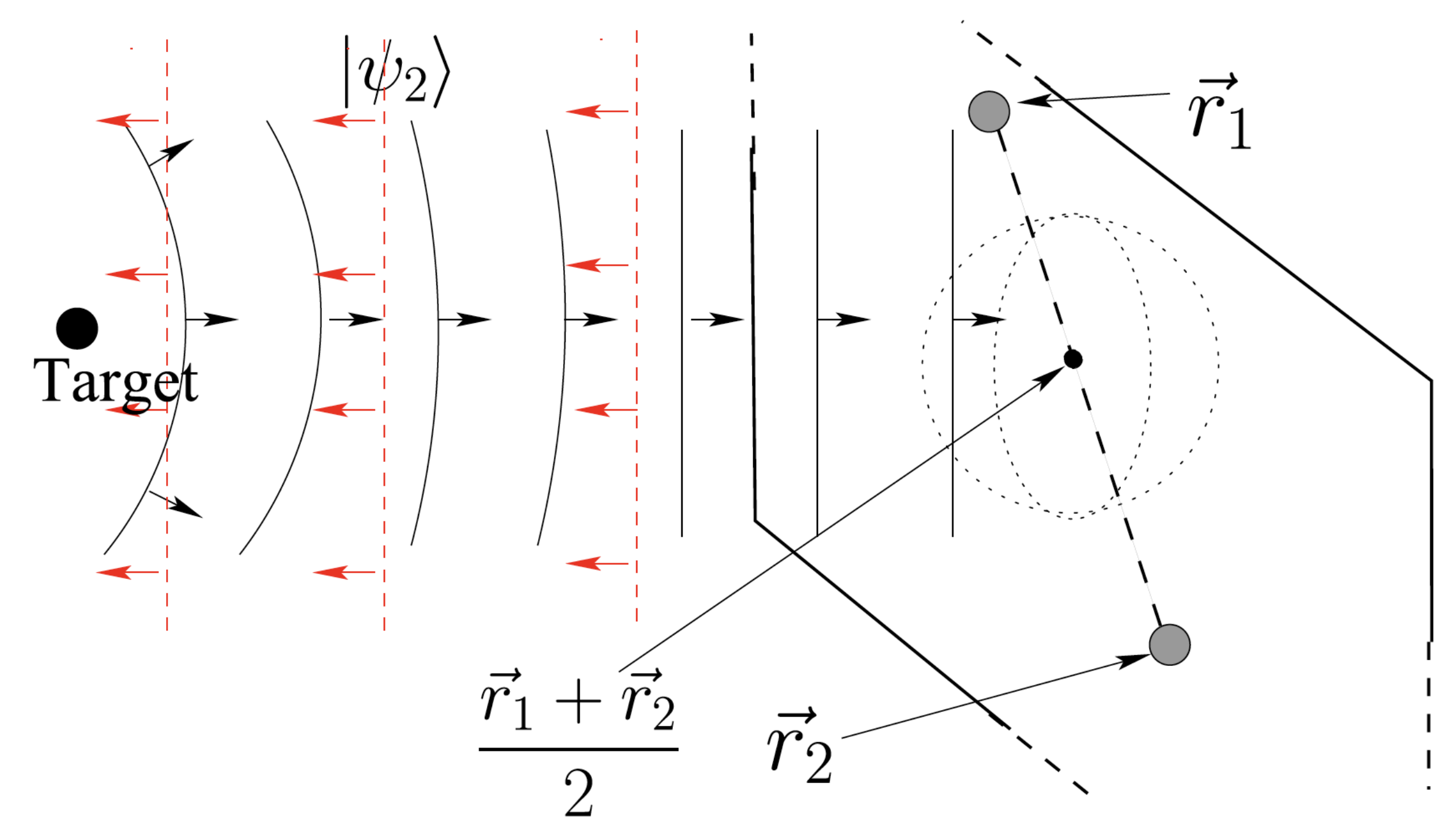}
\caption{Quantum radar setup (two-photon scenario): A target, resembling a point source, reflects the momentum-entangled state $|\psi_2\rangle$ of two incoming photons, depicted as dashed lines. In the far field, these photons reach a screen. The average time of arrival, although not shown in the illustration, provides the longitudinal distance measurement, while the average positions of the two photons' transverse arrivals, $\vec r_1$ and $\vec r_2$, yield the transverse location of the object, indicated by the dashed line. The uncertainty sphere, represented by the dotted line, experiences a reduction by a factor of $N^{3/2}$ compared to what would be obtained with N-independent photons sharing the same spatiotemporal bandwidth. Here $N=2$ \cite{MacconeRen2019}.}
\label{fig:MacconeRen}
\end{center}
\end{figure}
 The entangled state in Maccone-Ren's protocol is an EPR-like state of the form
\begin{align}
|\psi_N\rangle \equiv\,\int\,d\omega\,d\vec{k}\,\psi(\omega,\vec{k})\left( a^\dag (\omega,\vec{k})\right)^N|0\rangle,
\label{entangled states Maccone Ren}
\end{align}
where $ a^\dag (\omega,\vec{k})$ is the creation operator of a photon with frequency $\omega$ and transverse moment $\vec{k}=\,(k_x,k_y)$; the propagation of each of these photons is along the $z$-direction and they have the same momenta $\hbar \,k$. The function $\psi(\omega,\vec{k})$ is the $N$-{\it photon structure function}. For $N=2$, $\psi(\omega,\vec{k})$ is the  {\it biphoton structure function} of this particular class of states. One also assumes the far-field approximation
\begin{align*}
 |\vec{k}_3|^2 =\,(k^2_x+\,k^2_y+k^2_z)^2\gg (k^2_x+\,k^2_y )^2=|\vec{k}|^2
 \end{align*}
 Then the expression for the joint probability of detecting the $N$ photons at times $t_j$ and transverse locations $\vec{r}_j$ can be re-written in the far field approximation 
\begin{widetext}
\begin{align}
p(\{t_j,\vec{r}_j)_{j=1,...,N}\})\propto \,\left|\widetilde{\psi}\left(N\left(\frac{\sum^N_{j=1} t_j}{N}- t_0\right),N\left(\frac{\sum^N_{j=1} \vec{r}_j}{N}-\,\vec{r}_p\right)\right)\right|^2.
\label{probability Maccone-Ren2}
\end{align}
\end{widetext}
This expression has several relevant consequences. First, it provides a method to determine the target range.
From the detection times one can extract the target range, which is given by the expression
\begin{align}
r_z=\,c\,\frac{\sum^N_{j=1}(t_j-t_0)}{2\,N},
\label{target range in Maccone-Ren}
\end{align}
where $c$ is the speed of light.
The transverse location of the target is similarly estimated to be given by the average transverse displacement relation,
\begin{align*}
\vec{r}=\,\frac{\sum^N_{j=1}\,\vec{r}_j}{N}.
\end{align*}

For the experiment where one individual, non-entangled photon states are sent to explore and detect the target, the probability of detecting a single photon at time $t$ and transverse position $\vec{r}$ is given by 
\begin{align}
p(t,\vec{r})\propto \left|\widetilde{\psi}(t,\vec{r})\right|^2.
\label{probability of one individual photons}
\end{align}
The transverse location is given by the expectation value of the corresponding coordinates. Since the variance operation of a sum is the sum of the variance, there is a direct comparison between the statistical properties of \eqref{probability Maccone-Ren2} and \eqref{probability of one individual photons}. In particular, there is a suppression factor of order $1/\sqrt{N}$ in the variance of \eqref{probability Maccone-Ren2} with respect to the variance of \eqref{probability of one individual photons}. As explained in the paper, including the expected arrival time and transversal displacement detection vector, one can see that entanglement imprints a reduction in the range detection error of order $\sqrt{N}$. 

There are several relevant difficulties in the practical implementation of the Maccone-Ren protocol for quantum radar. First of all, the model used in \cite{MacconeRen2019} for the transfer function is not a realistic one, since the transfer function used is not the exact solution for the gauge potential. This flaw has been overcome in a new analysis in \cite{MacconeZhengRen}, where the Helmholtz equation for the potential is analyzed in detail.

Of more practical relevance is the inherent difficulty in the generation of the entangled states in Eq. \eqref{entangled states Maccone Ren}. A possible partial solution to this important problem is to use partially entangled states.
Second, the randomness in the distribution and arrival in time for the $N$ entangled photons implies an infinite time detection and infinite size of the detector. A partial solution for this issue was already anticipated by Maccone and Ref. \cite{MacconeRen2019}. It is shown that, although the enhancement in precision for the measurement of the range and transverse displacement is not as high as when using the maximally entangled states Eq. \eqref{entangled states Maccone Ren}, the use of partially entangled states
\begin{widetext}
\begin{align}
|\phi\rangle :=\,\int\,d\omega\,d\vec{k}\,\prod_j \,d\omega_j\,d\vec{k}_j \,\psi(\omega,\vec{k})\,\gamma(\omega_j)\,\xi(\vec{k}_j)\,a^\dag (\omega +\omega_j,\vec{k}+\vec{k}_j)|0\rangle .
\label{partially entangled states}
\end{align}
\end{widetext}
allows for a finite time of detection and a finite size of the detection screen. Furthermore, the states \eqref{partially entangled states} are easier to produce than the maximally entangled states \eqref{entangled states Maccone Ren} using SPDC methods, at least for states associated with two entangled photons \cite{Liu et al. 2009, Pires Exter 2009, Yun et al. 2012}.

The third issue in Maccone-Ren's quantum radar protocol is related with the effect of thermal noise, since the entangled states used in the protocol are very sensitive to noise. Indeed, usually protocols in quantum metrology are very sensitive to noise. In the case of Maccone-Ren quantum radar protocol, the lost of one of the $N$ entangled photons renders the other $N-1$ useless, since their detection is produced at random times and transverse locations. It has been suggested several strategies to solve this problem. Two of such suggestions are discussed in \cite{MacconeRen2019}. The first is to use the partially entangled states \eqref{partially entangled states}. Such states are more robust against noise. The second is suggested by the protocols discussed in \cite{GiovannettiLloydMaccone2002} and involve nested systems of entangled states. Both strategies reduce the effect of noise at the price of a reduction on the enhancement in precision by using entangled states for ranging detection.

\section{Conclusions and perspectives}

Quantum radar and quantum LiDAR, as quantum technologies, hold the promise of making a substantial impact on radar and scanning applications. Over the past few years, protocols like quantum illumination and related techniques have spurred both experimental efforts and theoretical developments. In this comprehensive review, we have presented various research directions and findings that, in our perspective, warrant cautious optimism regarding the potential for practical applications.

Nonetheless, the journey towards the realization of quantum radar and LiDAR is fraught with numerous challenges. At optical frequencies, the ubiquitous high-noise environment characteristic of typical atmospheric conditions isn't generally met. Furthermore, issues like signal fading pose additional constraints on the exploitation of sensitivity advantages in realistic scenarios \cite{ZhuangZhangShapiro}. In the microwave frequency domain, there are two primary hurdles to overcome in quantum radar development. Firstly, the requirement for extremely low signal intensities to benefit from quantum illumination, as opposed to coherent light, presents a formidable challenge. Secondly, achieving the ultra-low temperatures needed to generate entanglement remains a significant obstacle. In the context of accuracy enhancement applications, the requisite integration times pose a major practical challenge.
One approach to address these challenges involves identifying specific conditions where quantum illumination may still offer advantages, such as scenarios with intense jamming, particularly applicable to optical frequencies.

In the case of other quantum radar/LiDAR protocols, intrinsic difficulties exist as well. The Maccone-Ren protocol, for instance, is highly sensitive to noise and losses. Strategies have been proposed to mitigate these issues, but they often come at the cost of reducing the protocol's accuracy benefits \cite{MacconeRen2019}. Further research in this line of investigation is likely necessary to fully understand the potential of this protocol.

Interferometric quantum radar, the first type of quantum radar protocol discussed in the literature, also faces intrinsic challenges. Losses and atmospheric absorption effects mean that quantum interferometric radars based on NOON-states do not surpass the standard quantum limit (as in Eq. \eqref{optical path sensitivity}), except in scenarios with very low attenuation levels. In realistic attenuation scenarios, these NOON-based systems can even perform worse than the standard quantum limit, a situation exacerbated as the parameter $N$ increases \cite{Gilbert Hamrick Weinstein2008}. This presents a potential obstacle to the widespread use of quantum interferometric radar based on NOON-states \cite{GilbertWeinstein2008}. While alternative state types have been proposed \cite{Gallego Torrome Bekhtil Knott 2021}, detailed analyses are still required to assess their practical viability.

\textbf{Acknowledgment} S.B. acknowledges funding by the Natural Sciences and Engineering Research Council of Canada (NSERC) through its Discovery Grant, funding and advisory support provided by Alberta Innovates (AI) through the Accelerating Innovations into CarE (AICE) -- Concepts Program, support from Alberta Innovates and NSERC through Advance Grant project, and Alliance Quantum Consortium. This project is funded [in part] by the Government of Canada. Ce projet est financé [en partie] par le gouvernement du Canada.


\begin{thebibliography}{22}



\bibitem{Degen Reinhard Cappellaro} C. L. Degen, F. Reinhard, P. Cappellaro, {\it Quantum sensing}, Rev. Mod. Phys. {\bf 89}, 035002 (2017).
\bibitem{Pirandola} Pirandola, S., Bardhan, B.R., Gehring, T. et al. {\it Advances in photonic quantum sensing}. Nature Photon 12, 724–733 (2018). https://doi.org/10.1038/s41566-018-0301-6

\bibitem{Zheng} Z. Zhang \& Q. Zhuang,  {\it Distributed quantum sensing},  Quantum Science And Technology. \textbf{6}, 043001 (2021,7), https://dx.doi.org/10.1088/2058-9565/abd4c3

\bibitem{Aslam} Aslam, N., Zhou, H., Urbach, E.K. et al., {\it Quantum sensors for biomedical applications}, Nat Rev Phys 5, 157–169 (2023). https://doi.org/10.1038/s42254-023-00558-3

\bibitem{Harris 2009} Harris Corporation, {\it Quantum sensors program}, AFRL-RI-RS-TR-2009-208 Final Technical Report, August (2009).

\bibitem{Mayte}  Mayte Y. Li-Gomez, Pablo Yepiz-Graciano, Taras Hrushevskyi, Omar Calderón-Losada, Erhan Saglamyurek, Dorilian Lopez-Mago, Vahid Salari, Trong Ngo, Alfred B. U'Ren, and Shabir Barzanjeh, Quantum enhanced probing of multilayered samples, Phys. Rev. Research 5, 023170 (2023).

\bibitem{Skolnik} M. I. Skolnik, {\it Introduction to Radar Systems}, Third Edition, McGraw Hill Education (2001).

\bibitem{Slepyan} G. Slepyan, S. Vlasenko, D. Mogilevtsev and A. Boag, "Quantum Radars and Lidars: Concepts, realizations, and perspectives," in IEEE Antennas and Propagation Magazine, vol. 64, no. 1, pp. 16-26, Feb. 2022, doi: 10.1109/MAP.2021.3089994.

\bibitem{WAILOKLAI201858}Wai-Lok Lai, W., Dérobert, X. \& Annan, P. A review of Ground Penetrating Radar application in civil engineering: A 30-year journey from Locating and Testing to Imaging and Diagnosis. { NDT and E International}. \textbf{96} pp. 58-78 (2018).

\bibitem{Marco Lanzagorta 2011} M. Lanzagorta, {\it Quantum Radar}, Synthesis Lectures on Quantum Computing n. 5,  Morgan ad  Claypool publishers (2011).


\bibitem{Shapiro2019} J. H. Shapiro, {\it The Quantum Illumination Story},  IEEE Aerospace and Electronic Systems Magazine {\bf 35} , Issue: 4 , April 1 (2020).

\bibitem{Gallego Torrome Bekhtil Knott 2021}R. Gallego Torrom\'e, N. B. Bekhti-Winkel, and P. Knott, Introduction to quantum radar, arXiv:2006.14238 [quant-ph].

\bibitem{MacconeRen2019} L. Maccone and C. Ren, {\it Quantum Radar}, Phys. Rev. Lett. {\bf 124}, 200503 (2020).

\bibitem{MacconeZhengRen} L. Maccone, Y. Zheng and C. Ren, {\it Gaussian beam quantum radar protocol},  	arXiv:2309.11834 [quant-ph].


\bibitem{PhysRevLett.124.150502}Xia, Y., Li, W., Clark, W., Hart, D., Zhuang, Q. \& Zhang, Z. Demonstration of a Reconfigurable Entangled Radio-Frequency Photonic Sensor Network. {\em Phys. Rev. Lett.}. \textbf{124}, 150502 (2020,4)

\bibitem{Caves} Carlton M. Caves, {\it Quantum limits on noise in linear amplifiers}, Phys. Rev. D {\bf 26}, 1817  (1982).

\bibitem{Barzanjeh et al.} S. Barzanjeh, S. Guha, C. Weedbrook, D. Vitale, J. H. Shapiro and S. Pirandola, {\it Microwave quantum illumination}, Phys. Rev. Lett. {\bf 114}, 080503 (2015).

\bibitem{Zhong2013}L. Zhong et al, {\it Squeezing with a flux-driven Josephson parametric amplifier}, New J. Phys. {\bf 15} 125013 (2013).


\bibitem{Abdo} Baleegh Abdo, Archana Kamal, and Michel Devoret, Nondegenerate three-wave mixing with the Josephson ring modulator, Phys. Rev. B {\bf 87}, 014508 (2013).

\bibitem{Planat2020} Luca Planat, Arpit Ranadive, Rémy Dassonneville, Javier Puertas Martínez, Sébastien Léger, Cécile Naud, Olivier Buisson, Wiebke Hasch-Guichard, Denis M. Basko, and Nicolas Roch, Photonic-Crystal Josephson Traveling-Wave Parametric Amplifier, Phys. Rev. X {\bf 10}, 021021 (2020).

\bibitem{EOM} Ho Eom, B., Day, P., LeDuc, H. et al. A wideband, low-noise superconducting amplifier with high dynamic range. Nature Phys {\bf 8}, 623–627 (2012).

\bibitem{Usman} Usman A. Javid, Jingwei Ling, Jeremy Staffa, Mingxiao Li, Yang He, and Qiang Lin, Ultrabroadband Entangled Photons on a Nanophotonic Chip, Phys. Rev. Lett. {\bf 127}, 183601 (2021).

\bibitem{Savanier} Marc Savanier, Ranjeet Kumar, and Shayan Mookherjea, "Photon pair generation from compact silicon microring resonators using microwatt-level pump powers," Opt. Express {\bf 24}, 3313-3328 (2016).

\bibitem{Afzal} Shirin Afzal, Tyler J Zimmerling, Mahdi Rizvandi, Majid Taghavi, Taras Hrushevskyi, Manpreet Kaur, Vien Van, Shabir Barzanjeh, Bright quantum photon sources from a topological Floquet resonance, arXiv:2308.11451 .

\bibitem{Manzoni}C Manzoni and G Cerullo, Design criteria for ultrafast optical parametric amplifiers, J. Opt. {\bf 18} 103501 (2016).





\bibitem{Helstrom1976} C. W. Helstrom, {\it Quantum detection and Estimation Theory}, New York, NY : Academic Press (1976).

\bibitem{Chernoff} H. Chernoff, {\it A measure of the asymptotic efficiency for tests of a hypothesis based on the sum of observations}, Ann. Math. Stat., {\bf 23}, 493 (1952).

\bibitem{Audenaert et al.} K. M. R. Audenaert et al., {\it The Quantum Chernoff Bound}, Phys. Rev. Lett., {\bf 98}, 160501 (2007).

\bibitem{Weedbrook12} Christian Weedbrook, Stefano Pirandola, Raúl García-Patrón, Nicolas J. Cerf, Timothy C. Ralph, Jeffrey H. Shapiro, and Seth Lloyd, {\it Gaussian quantum information
}, Rev. Mod. Phys. {\bf 84}, 621 (2012).

\bibitem{Lloyd2008} S. Lloyd, {\it Enhanced Sensitivity of Photodetection via Quantum Illumination}, Sicence {\bf 321}, 1463 (2008).

\bibitem{Tan} S. -H. Tan et al., {\it Quantum Illumination with Gaussian States}, Phys. Rev. Lett. {\bf 101}, 253601 (2008).



\bibitem{ShapiroLloyd} J. H. Shapiro and S. Lloyd, {\it Quantum illumination versus coherent-state target detection}, New Journal of Physics {\bf 11} 063045 (2009).

\bibitem{Pirandola Lloyd} S. Pirandola and S. Lloyd, {\it Computable bounds for the discrimination of Gaussian states}, Phys. Rev. A {\bf 78}, 012331, (2008).

\bibitem{Di Candia et al.} R. Di Candia, H. Yi\u{g}itler, G. S. Paraoanu, R. J\"antti, {\it Two-way covert microwave quantum communication}, arXiv:2004.07192v1 [quant-ph].


\bibitem{de Palma Borregaard} G. De Palma and J. Borregaard, {\it Minimum error probability of quantum illumination,}
Phys. Rev. A {\bf 98}, 012101 (2018).

\bibitem{Guha Erkmen 2009} S. Guha and B. I. Erkmen, {\it Gaussian-state quantum illumination receivers for target detection}, Phys. Rev. A {\bf 80}, 052310 (2009).


\bibitem{Sanz et al.} M. Sanz, U. Las Heras, J. J. Garcia-Ripoll, E. Solano, R. Di Candia, {\it Quantum Estimation Methods for Quantum Illumination},  	Phys. Rev. Lett. {\bf 118}, 070803 (2017).

\bibitem{Zhang et al. 2015} Z. Zhang et al., {\it Entanglement-enhanced sensing in a lossy and noisy environment}, Phys. Rev. Lett. {\bf 114}, 110506 (2015).

\bibitem{Zhuang Zhang Shapiro b} Q. Zhuang, Z. Zhang, and J. H. Shapiro, {\it Optimum Mixed-State Discrimination for Noisy Entanglement-Enhanced Sensing}, Phys. Rev. Lett. {\bf 118}, 040801  (2017).

\bibitem{Reichert et al. 2023} M. Reichert, Q. Zhuang, J. H. Shapiro, R. Di Candia
{\it Quantum Illumination with a Hetero-Homodyne Receiver and Sequential Detection}, arXiv:2303.18207 [quant-ph].


\bibitem{Wald 1945} A. Wald, {\it Sequential tests of statistical hypotheses}, Ann.
Math. Stat. {\bf 16}, 117 (1945).

\bibitem{Karsa Pirandola 2021} A Karsa, S Pirandola, {\it Energetic considerations in quantum target ranging}, 2021 IEEE Radar Conference (RadarConf21), 1-4.

\bibitem{Jonsson et al.} R. Jonsson, R. Di Candia, M. Ankel, A. Ström, G. Johansson, {\it A comparison between quantum and classical noise radar sources}, 2020 IEEE Radar Conference (RadarConf20), Florence, Italy, {\bf 2020}, pp. 1-6.


\bibitem{ZhuangZhangShapiro} Q. Zhuang, Z. Zhang, and J. H. Shapiro, {\it Quantum illumination for enhanced detection of Rayleigh-fading targets}, Phys. Rev. A {\bf 96}, 020302(R) (2017).


\bibitem{Barzanjeh et al.2019} S. Barzanjeh, S. Pirandola, D. Vitali and J. M. Fink, {\it Microwave Quantum Illumination using a digital receiver}, Science Advances Vol. {\bf 6}, no. 19, eabb0451 (2020).

\bibitem{Luong et al. 2019} D. Luong, C. W. Sandbo Chang, A. M. Vadiraj, A. Damini, C. M. Wilson and B. Balaji, {\it Receiver Operating Characteristics for a Prototype Quantum Two-Mode Squeezing Radar},  IEEE Transactions on Aerospace and Electronic Systems, Volume: {\bf 56}, Issue: 3, June (2020).

\bibitem{Heshami et al. 2016} Heshami K, England DG, Humphreys PC, Bustard PJ, Acosta VM, Nunn J, Sussman BJ (November 2016). "Quantum memories: emerging applications and recent advances"J Mod Opt. 2016 Nov 12; {\bf 63}
 (20): 2005–2028. 
 
\bibitem{PhysRevLett.109.130503}Barzanjeh, S., Abdi, M., Milburn, G., Tombesi, P. \& Vitali, D. Reversible Optical-to-Microwave Quantum Interface. {\em Phys. Rev. Lett.}. \textbf{109}, 130503 (2012,9), https://link.aps.org/doi/10.1103/PhysRevLett.109.130503

\bibitem{PhysRevA.84.042342}Barzanjeh, S., Vitali, D., Tombesi, P. \& Milburn, G. Entangling optical and microwave cavity modes by means of a nanomechanical resonator. {\em Phys. Rev. A}. \textbf{84}, 042342 (2011,10), https://link.aps.org/doi/10.1103/PhysRevA.84.042342

\bibitem{Lauk_2020}Lauk, N., Sinclair, N., Barzanjeh, S., Covey, J., Saffman, M., Spiropulu, M. \& Simon, C. Perspectives on quantum transduction. {\em Quantum Science And Technology}. \textbf{5}, 020501 (2020,3), https://dx.doi.org/10.1088/2058-9565/ab788a

\bibitem{Barzanjeh2019}Barzanjeh, S., Redchenko, E., Peruzzo, M., Wulf, M., Lewis, D., Arnold, G. \& Fink, J. Stationary entangled radiation from micromechanical motion. {\em Nature}. \textbf{570}, 480-483 (2019,6), https://doi.org/10.1038/s41586-019-1320-2

\bibitem{Barzanjeh2022}Barzanjeh, S., Xuereb, A., Gröblacher, S., Paternostro, M., Regal, C. \& Weig, E. Optomechanics for quantum technologies. {\em Nature Physics}. \textbf{18}, 15-24 (2022,1), https://doi.org/10.1038/s41567-021-01402-0

\bibitem{Arnold2020}Arnold, G., Wulf, M., Barzanjeh, S., Redchenko, E., Rueda, A., Hease, W., Hassani, F. \& Fink, J. Converting microwave and telecom photons with a silicon photonic nanomechanical interface. {\em Nature Communications}. \textbf{11}, 4460 (2020,9), https://doi.org/10.1038/s41467-020-18269-z


\bibitem{Chang et al. 2019}   C. W. Sandbo Chang, A. M. Vadiraj, J. Bourassa, B. Balaji, and C. M. Wilson, {\it Quantum enhanced noise radar}, Appl. Phys. Lett. {\bf 114}, 112601 (2019).

\bibitem {Assouly et al. 2023} R. Assouly, R. Dassonneville, T. Peronnin, A. Bienfait and B. Huard, {\it Quantum advantage in microwave quantum radar}, Nature Physics (2023).


\bibitem{Livreri et al. 2022} P. Livreri, et al. {Microwave quantum radar using a Josephson
traveling wave parametric amplifier. In IEEE Radar Conference}, (RadarConf22) 1–5 (IEEE, 2022).


\bibitem{Esposito2022} M. Esposito et al., {\it Observation of Two-Mode Squeezing in a Traveling Wave Parametric Amplifier}, Phys. Rev. Lett. {\bf 128}, 153603 (2022).

\bibitem{Qiu et al. 2023} J. Y. Qiu et al. {\it Broadband squeezed microwaves and amplification with a Josephson travelling-wave parametric amplifier}, Nature Physics volume {\bf 19}, 706-713 (2023).

\bibitem{Lopaeva et al.} E. D. Lopaeva et al., {\it Experimental realization of quantum illumination}, Phys. Rev. Lett. {\bf 111}, 010501 (2013).



\bibitem{England Balaji Sussman 20019} D. G. England, B. Balaji, and B. J. Sussman1, {\it Quantum-enhanced standoff detection using correlated photon pairs}, Phys. Rev. A {\bf 99}, 023828 (2019).

\bibitem{Zhao et al. 2022} Zhao et al., {\it Light detection and ranging with entangled photons}, Opt. Express 30, 3675-3683 (2022)

\bibitem{Frick McMillan Rarity 2020} S. Frick, A. McMillan, J Rarity, {\it Quantum Rangefinding}, Opt. Express 28, 37118-37128 (2020).

\bibitem{Liu2019}Liu, H. et al. Enhancing lidar performance metrics using continuous-wave photon-pair sources. Optica 6, 1349 (2019).

\bibitem{Liu2020} Liu, H., Balaji, B. and Helmy, A. S. Target detection aided by quantum temporal correlations: Theoretical analysis and experimental validation. IEEE Trans. Aerosp. Electron. Syst. {\bf 56}, 3529 (2020).

\bibitem{He2020}He, H. et al. {\it Non-classical semiconductor photon sources enhancing the performance of classical target detection systems}, J. Lightwave Technol. {\bf 38}, 4540 (2020).

\bibitem{Blakey}Blakey, P.S., Liu, H., Papangelakis, G. et al. Quantum and non-local effects offer over 40 dB noise resilience advantage towards quantum lidar. Nat Commun {\bf 13}, 5633 (2022).

\bibitem{Liu} Liu, H., Qin, C., Papangelakis, G. et al. Compact all-fiber quantum-inspired LiDAR with over 100 dB noise rejection and single photon sensitivity. Nat Commun 14, 5344 (2023).


\bibitem{Ricardo2020b} R. Gallego Torrom\'e {\it Quantum illumination with multiple entangled photons},  Advanced Quantum Technologies, Volume {\bf 4}, Issue 11, DOI: 10.1002/qute.202100101.

\bibitem{Ricardo 2023} R. Gallego Torrom\'e, {\it Non-Gaussian Quantum Illumination with three modes}, arXiv:2305.10458 [quant-ph].

\bibitem{Su-Yong Lee et al. 2022} Su-Yong Lee et al.,  {\it Bound for Gaussian-state Quantum Illumination using direct photon measurements},  arXiv:2210.01471v2 [quant-ph].

\bibitem{Nair Gu 2020} R. Nair and M. Gu, {\it Fundamental limits of quantum illumination}, Optica, Vol. 7, 771 (2020)

\bibitem{Su-Yong Lee et al. 2021} S.-Y. Lee, Y. S. Ihn, and Z. Kim, {\it Quantum illumination via quantum-enhanced sensing}, Phys. Rev. A 103, 012411 (2021).

\bibitem{Noh Lee Lee} Changsuk Noh, Changhyoup Lee, Su-Yong Lee, {\it Quantum illumination with non-Gaussian states: Bounds on the minimum error probability using quantum Fisher information}, arXiv:2110.06891 [quant-ph].

\bibitem{Jung and Park 2022} E. Jung and D. Park, {\it Quantum Illumination with three modes Gaussian State}, Quantum Information Processing volume {\bf 21}, Article number: 71 (2022).

\bibitem{Daum et al. 2o21} F. Daum, J. Huang and A. Noushin, {\it Optimal quantum radar vs. optimal classical radar with full polarization antennas},  Proceedings of IEEE Radar Conference, Atlanta, May (2021).

\bibitem{Djordjevic 2022a} I. B. Djordjevic, {\it On Entanglement-Assisted Joint Monostatic-Bistatic Radars}, Entropy {\bf 24}, 756 (2022).

\bibitem{Djordjevic 2022b} I. B. Djordjevic, {\it On Entanglement-Assisted Multistatic Radar Techniques}, Entropy {\bf 24}, 990 (2022).

\bibitem{Karsa et al. 2020} A. Karsa, G. Spedalieri, Q. Zhuang and S. Pirandola, {\it Quantum Illumination with a generic Gaussian source}, Phys. Rev. Research {\bf 2}, 023414 (2020).

\bibitem{GiovannettiLloydMaccone} V. Giovannetti, S. Lloyd and L. Maccone, {\it Quantum-Enhanced Measurements: Beating the Standard Quantum Limit}, Science {\bf 306}, 1330 (2004).

\bibitem{PhysRevLett.106.090504}Pirandola, S. Quantum Reading of a Classical Digital Memory. {\em Phys. Rev. Lett.}. \textbf{106}, 090504 (2011,3), https://link.aps.org/doi/10.1103/PhysRevLett.106.090504


\bibitem{Zhuang Shapiro 2022} Q. Zhuang and J. H. Shapiro, {\it Ultimate Accurary Limit of Quantum Pulse-Compression Ranging}, Phys. Rev. Lett. {\bf 128}, 010501 (2022).

 \bibitem{Reichert et al. 2022}   M. Reichert, R. Di Candia, M. Z. Win and M. Sanz, {\it Quantum-enhanced Doppler lidar},
npj Quantum Information volume 8, Article number: 147 (2022).


 \bibitem{Huang et al. 2021} Z. Huang, C. Lupo and P. Kock, {\it Quantum-Limited Estimation of Range and Velocity}, Phys. Rev. X Quantum {\bf 2}, 030303 (2021).

 
\bibitem{Zhong et al. 2023} W Zhong et al.  {\it Relation between quantum illumination and quantum parameter estimation}
arXiv preprint arXiv:2308.07150.


\bibitem{GiovannettiLloydMaccone2001}  V. Giovannetti, S. Lloyd and L. Maccone, {\it Quantum enhanced positioning and clock synchronization}, Nature {\bf 412}, 417 (2001).



\bibitem{Boto et al.} A. N. Boto et al., {\it Quantum interferometric optical lithography: Exploiting Entanglement to beat the diffraction limit}, Phys. Rev. Lett. {\bf 85}, 2733 (2000).

\bibitem{GilbertWeinstein2008} G. Gilbert and Y. S. Weinstein, {\it Aspects of practical remote quantum sensing}, J. of Mod. Optics {\bf 55} 10 (2008).

\bibitem{MargulisLevitin} N. Margulus and L. B. Levitin, {\it The maximal speed of dynamical evolution}, Physica D {\bf 120}, 188 (1998).

\bibitem{GilbertHamrick} G. Gilbert and M. Hamrick,  MITRE Technical Report, (2000), {\it Practical Quantum Cryptography: A Comprehensive Analysis (Part One)}, http://arXiv.org/abs/quant-ph/0009027.

\bibitem{Gilbert Hamrick Weinstein2008} G. Gilbert, M. Hamrick, and Y. S. Weinstein {\it Use of maximally entangled N-photon states for practical quantum interferometry}, Vol. {\bf 25}, Issue 8, 1336 (2008).

\bibitem{Dowling} J. P. Dowling, {\it Correlated input-port, matter-wave interferometer: Quantum-noise limits to the atom-laser gyroscope}, Phys. Rev. A {\bf 57}, 4736 (1998).

\bibitem{Yurke et al. 1986} B. Yurke, S. L. McCall and J R. Klauder, {\it SU(2) and SU(1,1) interferometers}
Phys. Rev. A {\bf 33}, 4033 (1986).

\bibitem{Liu et al. 2009} W. Liu, P-x. Chen, C. Z. Li and J-M. Yuan, {\it Preparation and identification of two-photon positively momentum correlated entangled states}, Phys. Rev. A {\bf 79}, 061802 (2009).

\bibitem{Pires Exter 2009} H. Di Lorenzo Pires and M. P. van Exter,{\it Observation of near field correlations in expontaneous parametric down conversion}, Phys. Rev. A {\bf 79}, 041801 (2009).

\bibitem{Yun et al. 2012} S. Yun et at., {\it  Generation of positively momentum correlated biphotons from spontaneous parametric down conversion}, Phys. Rev. A {\bf 86}, 023852 (2012).


\bibitem{GiovannettiLloydMaccone2002}  V. Giovannetti, S. Lloyd and L. Maccone, {\it Positioning and clock synchronization through entanglement}, Phys. Rev. A {\bf 65}, 022309 (2002).




\end{thebibliography}
\end{document}